 \documentclass[preprint]{aastex}

\shorttitle{Speckle Measures with DSSI. VI.}
\shortauthors{Horch et al.}

\begin{document}
\newcommand{\beq}{\begin{equation}}
\newcommand{\eeq}{\end{equation}}
\newcommand{\ea}{{\it et al.\ }}


\title{Observations of Binary Stars with the Differential Speckle
Survey Instrument. VI. Measures During 2014 at the Discovery Channel Telescope
}

\author{Elliott P. Horch\altaffilmark{1}}
\affil{Department of Physics, 
Southern Connecticut State University,
501 Crescent Street, New Haven, CT 06515}
\email{horche2@southernct.edu}

\author{Gerard T. van Belle}
\affil{Lowell Observatory, 1400 West Mars Hill Road, 
Flagstaff, AZ 86001}
\email{gerard@lowell.edu}

\author{James W. Davidson, Jr.\altaffilmark{2}, Lindsay A. Ciastko}
%\author{James W. Davidson, Jr.\altaffilmark{2}}
\affil{Department of Physics, Albion College,
611 E. Porter Street,
Albion, MI 49224}
\email{james.w.davidson.jr@gmail.com, lac13@albion.edu}
%\email{james.w.davidson.jr@gmail.com}

\author{Mark E. Everett}
\affil{National Optical Astronomy Observatory, 950 North Cherry Street,
Tucson, AZ 87719}
\email{everett@noao.edu}

\author{Karen S. Bjorkman}
\affil{Department of Physics and Astronomy, University of Toledo,
Toledo, OH 43606}
\email{karen.bjorkman@utoledo.edu}

\altaffiltext{1}{Adjunct Astronomer, Lowell Observatory.}
\altaffiltext{2}{Current Address: Department of Astronomy, University
of Virginia, P.O.\ Box 400325, Charlottesville, VA 22904-4325.}

%\newpage

\begin{abstract}

We present the results of 938 speckle measures of double stars and
suspected double stars drawn
mainly from
the {\it Hipparcos} Catalogue, as well as 208 observations where no companion
was noted.
One hundred fourteen pairs have been resolved for the first time.
The data were obtained during four observing runs in 2014
using the Differential Speckle Survey Instrument
(DSSI) at Lowell Observatory's Discovery Channel Telescope. 
The measurement precision
obtained when comparing to ephemeris positions of binaries with very
well-known orbits is generally less than 2 mas in separation and 
0.5 degrees in position angle. Differential photometry is found 
to have internal precision of approximately 0.1 magnitudes and to be
in very good agreement with {\it Hipparcos} measures in cases where the comparison
is most relevant. We also 
estimate the detection limit in the cases where no companion was found. 
Visual orbital elements are derived for 6 systems.

\end{abstract}


\section{Introduction}

Observations of binary stars have traditionally played an important role
in our knowledge of stellar structure and evolution, and this remains 
a vital field primarily because of the increased resolution that can be
obtained for imaging of binaries as well as the precision
in radial velocity and metallicity measurements that can be 
derived from stellar 
spectra. Over the last two decades, this has 
resulted in a dramatic increase in the number of systems 
that can be studied with multiple techniques. 
A few examples of resolving 
close binaries and deriving precise relative astrometry
include the work of Hummel \ea (1995, 1998) on spectroscopic
binaries with both the
Mark III and Navy Optical Interferometers, 
a similar program at the Palomar Testbed Interferometer 
(Muterspaugh \ea 2010a and references therein),
the comprehensive study of Torres (2009) on $\alpha$ Aurigae, 
the extremely productive
speckle program of Mason, Hartkopf, and their collaborators (Mason \ea 2013),
the speckle program of Tokovinin and his collaborators 
in the Southern Hemisphere (Tokovinin \ea
2015 and references therein), the work of Balega \ea (2013), 
and our own work at the WIYN and Gemini telescopes (Horch \ea 2011a, 2011b, 
2012, 2015).
In addition, both the {\it Hipparcos} Catalogue and the Geneva-Copenhagen 
spectroscopic survey give a ready list of binaries, some of which will
become important if future observations can be used to
establish their orbital parameters.
This is a particularly 
favorable situation for the derivation of stellar masses of sufficient
quality to make progress on the understanding of the dependence of mass
on parameters such as metallicity and age. In the best case of 
a resolved binary that also is a double-lined spectroscopic system,
a distance measure independent of {\it Hipparcos} can be derived.

The structure and physical 
parameters of binary and multiple stars are informative from a 
statistical point of view regarding star formation theories, as shown
most recently in papers such as Raghavan \ea (2010), 
Tokovinin (2014a, 2014b), and Riddle \ea (2015).
In addition, it has been realized in the last couple of years that 
a significant number of stars that host exoplanet systems also have a 
stellar companion (e.g.\ Lillo-Box \ea 2014, Kane \ea 2014, Horch \ea 2014), 
although there is evidence 
that the presence of 
stellar secondaries is suppressed relative to the field at smaller 
separations (Wang \ea 2014a, 2014b).
High-precision astrometric measurements of binaries 
and multiple stars that are relatively close to the Sun are a key 
prerequisite to exploring these issues.

With these facts in mind, we have started a new program of speckle 
observations of double stars at Lowell Observatory's 
Discovery Channel Telescope (DCT), a 4.3-meter
telescope recently commissioned by the observatory and its partner
institutions. The observing program is largely an extension of work previously
done by two of us (E.H. and M.E.) at the WIYN 3.5-m telescope at Kitt Peak,
together with other collaborators. However, we also have an interest in 
the current program to identify and study nearby and young binaries, whereas the WIYN
program focused more on differentiating between the thin and thick disk samples
of binaries. The primary sources of potential targets remain the {\it Hipparcos}
and Geneva-Copenhagen Catalogues. We detail here the first year of observations
on the program, which used the Differential Speckle Survey Instrument (DSSI).

\section{Observations}

The DSSI was on the telescope on four occasions during 2014: two nights
in March, two in June, eight nights from 30 September to 7 October, 
and four more in November
for a total of 16 nights, of which approximately five were used 
for binary star observations reported here. The instrument is described 
in Horch \ea (2009), and the subsequent upgrade to the use of 
electron-multiplying CCD cameras
instead of the original low-noise, large format CCD chips is described 
in Horch \ea (2011a).
A number of other observational programs were executed during the four runs, 
including high-resolution
imaging of minor planets and follow-up observations for the 
{\it Kepler} satellite mission; those will be discussed in
future papers. In between the spring and 
fall runs, the instrument was used at the Gemini North telescope, thus 
it was disassembled and packed for shipping
both to and from Hawaii. The instrument was designed with portability 
in mind, so this procedure is relatively straight-forward, but it does
mean that the plate scale calibrations for the later two runs at the DCT can be 
expected to be slightly different than for the first two runs.
The seeing for the observations at the DCT presented 
here ranged from 
0.6 to 1.2 arc seconds, with an average value of 0.8 arc seconds.

All observations taken here consisted of sets of 
1000 short exposure images of the
target star, with each image being 40 ms in duration. 
In the case of some fainter 
objects, two or three such sets were taken 
and co-added in the analysis phase. 
A 128$\times$128 subarray was used for all observations 
discussed here; this gave DSSI a
field of view of about 2.4$\times$2.4 arc seconds at the DCT. 
Prior to observing, the target
list was put into groups of two to several stars 
that had sky positions within a few
degrees of one another. These were then generally 
observed in order of increasing 
right ascension starting when the first target was 
near the meridian. An unresolved source
taken from the Bright Star Catalogue was used as a 
calibration object for the 
purposes of determining the speckle transfer function for each group of science 
targets, and placed within the group based on its right ascension. In this way, 
the zenith distance of all observations in the group was kept to a minimum. 

Figure 1 shows two representations of the data 
set for the systems where a companion 
was detected; in Figure 1(a), the data are plotted 
in terms of the magnitude difference
obtained as a function of the measures separation, 
and in Figure 1(b), the magnitude
differences are plotted as a function of the 
total (system) magnitude appearing
%in the SIMBAD online database\footnotemark. 
%\footnotetext{{\tt http://simbad.u-strasbg.fr/simbad}}
in the {\it Hipparcos} Catalogue (ESA 1997).
These figures are comparable to previous
work with the instrument at the WIYN 3.5-m 
telescope (e.g.\ Horch \ea 2011a), and are effectively 
semi-log plots since the $y$-axis is proportional to the log
of the brightness ratio. In particular, the envelope
of points in Figure 1(a) shows that, starting 
at small separations, the 
sensitivity to the magnitude
difference of companions rises sharply, but a 
``knee'' appears at a separation of 
approximately 0.1 arc seconds ($\Delta m$ of 3-4 magnitudes), 
and at larger separations the sensitivity 
continues to rise, albeit at a more modest 
pace. 
%This is actually the result of the fact 
%that the plot is in effect a semi-log plot since 
%the $y$-axis gives the magnitude
%difference. If plotted as a log-log plot, the 
%envelope of points would be close to
%linear, crossing the $x$-axis at a separation 
%value of approximately $\log(0.01)=-2$
%and having a value of approximately 5 magnitudes 
%at separation $\log(1.0)=0$.
One can also note that there are some systems 
with measured separations below the 
nominal diffraction limit at the DCT of about 
0.04 arc seconds (at 692 nm). This
is possible due to the fact that the DSSI 
takes images in two filters simultaneously.
As discussed in Horch \ea 2011b (Paper III of this series), 
the two-color analysis permits us
to distinguish between atmospheric dispersion 
and the presence of a companion
below the diffraction limit.
Generally, to minimize the effect of atmospheric dispersion,
we observed the targets at an airmass of less than 2.

\section{Data Reduction and Analysis}

\subsection{Determination of the Pixel Scale and Orientation}

In our work at the WIYN telescope with DSSI, 
we constructed a slit mask for use in the determination
of the pixel scale and orientation at that 
telescope. We were able to attach the mask
to the tertiary mirror baffle support and, with 
the help of WIYN staff, were able to
easily mount and remove the mask with minimal 
loss of observing time. This essentially 
follows the example of the CHARA and USNO 
speckle programs, and is generally considered
the most reliable method for precise scale 
and orientation calibration for speckle
measures. However, at the DCT, we do not yet 
have such a system in place since the tertiary mirror is
located inside the instrument cube and therefore inaccessible. As a result, 
we were forced in 2014 to rely on a group
of calibration binaries during 
each run. This limits the precision of our 
measures somewhat, especially at the 
larger separations that we report here, since 
the scale in arc seconds per pixel is 
multiplied by the separation in pixels to 
obtain the final separation.
However, we will show that there is no 
evidence for a systematic offset in
separations we obtain. In the future, we plan to develop a system by which
a calibration mask can be placed inside the DSSI camera 
itself. This would allow us to make a robust scale determination
from first principles regardless of the telescope with which DSSI is used,
and will be important to ensure 
the long-term consistency of our speckle measures.

On each run, we selected a small number of binaries 
with extremely well-known orbits
with which to measure the scale. We required that the 
the orbits used include recent data and that the ephemeris
uncertainties in position angle and separation be less than or
equal to 0.5 degrees and 1 mas respectively. 
Table 1 shows the 
final scale in mas per pixel and the offset angle between celestial
coordinates and the pixel axes that were applied on each run.
Because we observed very few objects in November for this project, we did
not have the opportunity to observe scale objects, so the results from 
the September/October run were applied.

Table 2 shows the objects used to obtain the results in Table 1 for
each run.
The columns give (1) the month of the run, (2) the WDS number of the binary
used, (3) the discoverer's designation, (4) the {\it Hipparcos} number, (5) the 
Besselian year of our observation, (6-9) the observed minus calculated residual
in position angle and separation for both the A and B cameras, and (10)
the reference for the orbit used to do the calculation.
Although we did not take data on scale calibration objects in November, we did
however measure the scale in another way, albeit at lower precision. We
took a sequence of 1-second exposures on a bright star, offsetting the telescope
in various directions in between the exposures. By measuring the 
centroid position of the star in each case and calculating the shift
between frames, we were able to compare this to the number of arc seconds
we had moved the telescope, and thereby determine both the scale and
offset angle. This method 
allowed us to conclude that the scale and 
offset angles in 
November were 
consistent with those derived for the September/October run. Therefore,
we used the latter to all of the November data.

Two things may be noted from Table 2. 
First, for each run shown, the average
residual is zero (or very near zero) for both position angle and separation
in the A camera, and for the position angle in the case of the B camera.
However, the average residual is not zero for the separation residuals for the
B camera. Specifically, for the five measures used in March, the average
residual is $-$0.26 mas, and for September/October, it is $-$0.56 mas.
This is because in this channel, DSSI is known to have a scale distortion
which is dependent on position angle that is related to the positioning of
one of the optical components in the optical train. 
More information about this distortion can be found in Horch \ea 2009 and
Horch \ea 2011a.
For the data presented
here, we assumed that the offset of this element was the same as what
we have measured at the WIYN telescope. If this parameter varied slightly
from the WIYN value,
it would be possible to obtain a non-zero residual for this purpose. By 
varying the scale, one can minimize, but not eliminate, an average
residual in this case.
This is not an issue for position angle or the scale in the A camera, which
does not have this problem.
Another way to check that the scale is not influenced by this is to 
compare results from the A and B camera for all observations; we 
show that these do not have a significant difference in Section 3.3.1. 
We conclude that the variations we see in Table 2 are likely 
dominated by random error.

Secondly, we may use the values in Table 2 to derive a simple estimate of
the measurement precision by calculating the standard deviation of each
type of residual in Columns 6-9, if we assume that the error in the 
ephemeris position is negligible. In this case we obtain 
$\sigma_{\theta,A} = 0.31 \pm 0.06^{\circ}$ and 
$\sigma_{\theta,B} = 0.26 \pm 0.05^{\circ}$ in position angle and
$\sigma_{\rho,A} = 1.7 \pm 0.4$ mas and 
$\sigma_{\rho,B} = 2.5 \pm 0.5$ mas in separation. Averaging, this 
indicates internal precision of $\sim$0.3 degrees in position angle and
$\sim$2 mas in separation.

\subsection{Speckle Data Reduction}

The reduction of science data followed the same sequence as described 
in previous papers in this series ({\it e.g.\ }Horch \ea 2011a, Horch \ea 2015).
The autocorrelation function is formed for each frame of the 
speckle exposure, and these are summed. Near-axis subplanes of the image
bispectrum are also calculated for each frame and summed.
Then, using the method of Meng \ea (1990), we form an estimate of the 
phase function for the object in the Fourier plane. We combine this
with the modulus of the power spectrum (derived from the autocorrelation
function deconvolved by the observation of a point source) 
to have a full estimate of the Fourier transform of the
object. This is low-pass filtered with a Gaussian function where the width
is chosen to give a full width at half maximum that is similar to that
of a diffraction-limited point spread function. Finally, to obtain a 
diffraction-limited reconstructed image, the result is inverse-transformed.
To give an indication of the typical quality of the results, 
reconstructed images for two objects, HDS 2143 and CHR 74, are shown in 
Figure 2. Orbits for each object are presented later in the paper.

The reconstructed image is then examined for companions visually. If a 
companion is noted, then the pixel location of the secondary peak is noted
and used as the starting position for a fitting program that performs 
a weighted least-squares fit to the object's power spectrum. In the case
of a successful fit, the measure has been added to our main list of
results. 
If, on the other hand, a companion is not noted visually, we then compute
a detection limit as a function of separation from the target. More about
this process is given in the Section 3.4.

\subsection{Double Star Measures}

In Table 3, we present our measures of double stars. The columns are
as follows:
(1) the Washington Double Star (WDS) number (Mason \ea 2001), which also
gives the right ascension and declination for the object in 2000.0 coordinates;
(2) the Aitken Double Star (ADS) Catalogue number,
or if none, the Bright Star Catalogue ({\it i.e.\ }Harvard Revised [HR]) number,
or if none, the Henry Draper Catalogue (HD) number,
or if none the Durchmusterung (DM) number of the object;
(3) the Discoverer Designation;
(4) the {\it Hipparcos} Catalogue number;
(5) the Besselian date of the observation;
(6) the position angle ($\theta$) of the
secondary star relative to the primary, with North through East defining the
positive sense of $\theta$; (7) the separation of the two stars ($\rho$), in
arc seconds; (8) the magnitude difference ($\Delta m$)
of the pair in the filter used;
(9) the center wavelength of the filter; and 
(10) the full width at half maximum of the filter transmission.
The measures have not been precessed from the dates shown.
One hundred fourteen objects in the table have no previous detection of the companion
in the 4th Catalogue of Interferometric Measures of Binary Stars (Hartkopf
\ea 2001b); we propose discoverer designations of
LSC (Lowell-Southern Connecticut) 2-115 here. (These numbers begin at 2
because LSC 1Aa1,Aa2 = HIP 103641 was discovered in previous work in collaboration
with Lowell [Horch \ea 2012].)

\subsubsection{Astrometric Accuracy and Precision}

Since we have a camera that takes data in two colors simultaneously, we 
can use the independent results in each channel to make a determination
regarding the intrinsic precision of our measures. In Figure 3, we show
the residuals obtained when subtracting the position angle and  
separation obtained in Channel B (880 nm) 
from those obtained in Channel A (692 nm).
Figure 3(a) shows the position angle result for all measures in Table 3; 
these show a negligible relative offset:
the average difference obtained is $0.04 \pm 0.11$ degrees, and 
the standard deviation obtained from all measures is $2.24 \pm 0.07$ degrees.
The latter is expected to increase as the separation decreases, since
the linear measurement uncertainty orthogonal to the separation subtends
a larger angle at smaller separations. If only measures with separations
larger than 0.1 arc seconds are included, then the average difference
remains consistent with zero, namely $-0.08 \pm 0.12$ degrees, and the
standard deviation decreases to $0.51 \pm 0.09$ degrees. 

In Figure 3(b-d), the separation differences are shown as a function of average
separation, average position angle, and average magnitude difference
respectively. From the entire data set, we obtain an average difference
between the channels of $-0.39 \pm 0.12$ mas with a standard deviation 
of $2.57 \pm 0.09$ mas. Such an offset might arise from the distortion
in Channel B of the instrument, which is known to depend on
position angle. However, when plotting the residuals as a function of 
position angle (Figure 3c), we see no clear trend. This is good evidence that
the distortion model applied is appropriate for DCT data; if this were not
the case, then the residuals would have a sinusoidal trend with two 
complete periods covered in the full range of 360 degrees in position angle.
On the other hand, when plotting the data as a function of average 
separation or magnitude difference, we see that there is a slight negative trend 
to the residuals for the measures reported below the diffraction limit (Figure 3b)
and for measures where the magnitude difference is large
(Figure 3d). 
For measures above the diffraction limit
and having average $\Delta m < 3$, we obtain an average A-B difference of
$=-0.07 \pm 0.11$ mas, with a standard deviation of $1.96 \pm 0.08$ mas.
Overall, we conclude that the scale and offset angles applied to the 
data yield no significant offsets between the channels for the bulk
of the measures presented, with the caveat that
below the diffraction limit and at large magnitude difference, there may
be small systematic offsets particularly in separation. 
It is also clear that at large magnitude
difference, the internal precision of the measures degrades.
The average difference in separation for measures below the diffraction
limit is $-1.20 \pm 0.44$ mas with standard deviation of $2.78 \pm 0.31$ mas,
and for measures with average magnitude difference greater than 3.0 we
have an average difference of $-0.95 \pm 0.34$ mas with 
a standard deviation of $3.60 \pm 0.24$ mas.
Thus, given the data at hand, the systematic offset in separation appears to be
on the order of perhaps 1 mas in both cases. This effect has so far not
been noticed at WIYN, and should be studied further
when more data at the DCT is available. 

Nonetheless, for the purposes 
of the present paper we have not corrected for this possible source
of error and we will use the intrinsic precision for all measures
derived from the first separation difference mentioned, $2.57 \pm 0.09$ mas.
Since this is a difference formed from two independent measures of
(assumed) equal precision, the standard deviation obtained will be 
$\sqrt{2}$ times the precision of an individual measure. So, we 
estimate that the intrinsic precision of separation measures is
$\sigma_{\rho} = (2.57 \pm 0.09)/ \sqrt(2) = 1.82 \pm 0.06$ mas.
This in turn suggests that the position angle precision should follow
the relation $\sigma_{\theta} = \arctan{(\sigma_{\rho} / \rho)}$, which
has value 0.2 degrees at a separation of approximately 0.5 arcsec; 
this is reasonably consistent with Figure 3(a).

There are a number of binaries observed that have orbits
of relatively high quality in the {\it Sixth Catalog of Visual Orbits 
of Binary Stars} (Hartkopf \ea 2001a)
that were not used in the determination of
the scale. We may use these to further judge the intrinsic accuracy and
precision of the measures in Table 3. A listing of these objects is
given in Table 4, together with the orbit information. 
We studied the position angle and separation values separately. 
To be used for the position angle study, we required that the orbit
have published uncertainties in the orbital parameters, and that 
for the epoch of our observation, the propagated uncertainty in 
position angle was less than or equal to 4 degrees.
For the separation study, we again required published uncertainties 
in the orbital parameters, and that the propagated uncertainty
in separation for the epoch of observation was less than or equal to 
4 mas. Figure 4 shows the observed minus ephemeris residuals
in both parameters plotted as a function of separation. In this case,
to make the plots clearer, we have averaged the observed result in
both channels prior to plotting. 
There are no noticeable offsets or trends in either coordinate. The
average residual in position angle is $0.16 \pm 0.21$ degrees, whereas
the average residual in separation is $-0.90 \pm 0.94$ mas with standard
deviation of $4.96 \pm 0.66$ mas. Many of the objects with
the largest residuals are also those with smallest separations. The
internal repeatability study suggests that the precision of these 
measures is probably not very different than those of larger separations.
It is therefore possible that the 
uncertainties for the orbital parameters for these objects may be 
slightly underestimated. If, on the other hand, we consider only those objects
with separations greater than 0.1 arcsec, then the average residual in
separation is
$0.47 \pm 0.81$ mas. Here, the
standard deviation of residuals is $2.91 \pm 0.57$ mas, with the average
ephemeris uncertainty being 1.96 mas. If we subtract the latter figure
from the former in quadrature, we can obtain another estimate of the
intrinsic measurement precision, where the result is $2.15 \pm 0.80$ mas,
which is consistent with the value obtained from comparing the A and
B channels.

\subsubsection{Photometric Precision}

We investigate the photometric precision of our measures by first 
defining a parameter which should 
approximately scale with the ratio of the measured separation
to the size of the isoplanatic
angle, as we have in previous papers in this series. Let this parameter,
which we call $q^{\prime}$, equal the seeing value for the observation
multiplied by the measured separation of the pair. (Since the seeing
should be roughly proportional to the inverse of the isoplanatic angle, the
desired ratio is obtained, up to a scaling factor.)

In Figure 5(a) we plot the difference between the magnitude difference
we obtain for each measure at 692 nm appearing in Table 3 and the value
of the magnitude difference appearing in the {\it Hipparcos} Catalogue
(ESA, 1997). This comparison is limited since the
center wavelength of the {\it Hipparcos} $H_{p}$ filter is significantly bluer
than either filter we have used. Nonetheless, the closest comparison we 
can make is with our 692-nm data. The plot shows what we have seen
in previous papers, namely that the residuals are flat out to a value 
of $q^{\prime} = 0.6$ arcsec$^{2}$ and that, at larger values, the
most reliable measures tend to have positive residuals. This is expected
since, as the separation becomes larger, there will be a loss of correlated
speckles between the primary and secondary speckle patterns, and therefore
a systematically large value of the magnitude difference will be obtained.
For this reason, if the measure has a $q^{\prime}$ value larger than 0.6
arcsec$^{2}$, then we have not included the magnitude difference in
Table 3.

Figure 5(b) shows the $\Delta H_{p}$ values with the lowest uncertainties
plotted as a function of the $\Delta m(692nm)$ value we have obtained.
We have eliminated objects which are giants, have indications in the {\it Hipparcos}
Catalogue of variability, and those with $B-V > 0.6$. The plot shows
a basically linear trend of slope near 1, as expected. The scatter about
the mean line is higher than {\it e.g.\ }Horch \ea 2011a, but this is
probably still due to the fact that we are comparing results in a 
blue filter with one in a red filter, whereas in the 2011 work, there were
many measures at 562 nm, a much better match to $H_{p}$. 
There are a number of examples in Table
3 of objects that were observed twice; the average difference 
in the magnitude difference obtained for these objects is 
$0.10 \pm 0.01$ in the 692-nm filter and $0.13 \pm 0.02$ in the 880-nm filter.
As these are subtractions of independent measures of the same 
intrinsic precision, we can estimate the internal repeatability of
our magnitude differences of $(0.10 + 0.13)/2 \sqrt{2} = 0.08 \pm 0.02$
magnitudes, very similar to results we have obtained at the WIYN 
and Gemini North telescopes.

Similarly, there are two objects, HIP 75312 and HIP 75695, that we
observed on three separate occasions. The average standard deviation 
of the magnitude difference in each filter for these two cases
is $0.068$ at 692 nm and $0.081$ at 880 nm.
Assuming this is not a significant difference, then we can average
all values regardless of filter
to obtain $0.075 \pm 0.012$, in excellent agreement
with the derived value from objects observed twice. 

\subsection{Nondetections}

There were 104 objects that were observed with both filters 
under good conditions and the
images obtained were judged to be of high quality, but did not show any
evidence of a companion. In these cases, we can estimate the limiting
magnitude difference as a function of separation from the central
star from the reconstructed images. Our method for doing so was 
described in Horch \ea 2011a, but briefly, we form annuli centered on
the peak of the central star of width 0.1 arc sec and separated by 0.1
arcsec. In each annulus, we locate all local maxima and local minima. 
We compute the average value and standard deviation of the local maxima.
We also compute the standard deviation of the absoluted value
of the local minima. Typically, this value is similar to that of the
local maxima, so we average these to obtain a more robust estimate
of the standard deviation of all extrema, and then consider the 
5-$\sigma$ detection limit to be the
average value plus five times this ``average'' standard deviation.

Figure 6 shows the result of this calculation for two objects: one
where no companion was detected, HIP 77986, which is the Be star
4 Her, and one where a companion was detected, namely HIP 113690 =
LSC 104. (The first measurement of this system appears in Table 3.)
In these plots, a detection-limit curve is traced out
using twelve concentric 
annuli, from 0.1 to 1.2 arcsec. The curve plotted
is a cubic spline interpolation, and it assumes that the curve
has limiting $\Delta m$ of zero at the diffraction limit. In the
case of LSC 104, we see a point representing a local maximum at separation
0.13 arcsec that is below the detection limit curve, indicating that this 
peak has greater than 5-$\sigma$ significance. In contrast, for 
HIP 77896, no peaks appear below the curve, indicating that no
second source was detected.

It is interesting to compare the shape of these detection limit curves
with Figure 1(a). The envelope of the points in Figure 1(a) has a
similar form, which indicates that, for high-quality observations, 
DSSI observations generally will not miss secondaries that fall within
the area defined by the typical detection limit curves, such as 
those seen in Figures 6(c,d).
Table 5 gives our final list of high-quality non-detections, with 
the detection limit obtained at 0.2 arcsec. This value is just above the
``knee'' in the curve, which gives a good reference point for the 
rest of the curve.

\section{Orbits for Six Systems}

The data presented here, together with other measures in the literature, 
provide the basis for the calculation of six orbits, two of which
(EGG 2Aa,Ab and YSC 133) 
are refinements, and four of which 
(HDS 2143, CHR 74, HDS 2947, and YSC 139)
are the first determination of 
orbital parameters for the system. We show the orbits in Figure 7,
and the final orbital parameters and other relevant data
are shown in Table 6. While none of these orbits may be considered
to be definitive at this time, this nonetheless identifies these 
systems as potentially useful for stellar astrophysics in the future, 
if further observations of high quality can be obtained. 
We discuss each system briefly below.

\subsection{EGG 2Aa,Ab}

This system has spectral type F2 from SIMBAD and 
parallax of $\pi = 10.26 \pm 2.92$ mas from van Leeuwen (2007).
A previous orbit was calculated by Cvetkovic (2008) with period
of 109 years and semi-major axis of 0.344 arcsec. However, our new
observation shows that the system has a much shorter period and smaller
semi-major axis. However, our orbit yields roughly the same 
total mass, with large
uncertainty. The system appears to have executed nearly a full orbit
since the first speckle observations taken of it in 1985 by McAlister
\ea (1987). More data will be needed to confirm this orbital trajectory,
but if correct, the smaller period makes the system much more attractive 
for future study.

\subsection{HDS 2143}

The composite spectrum of this pair is given in the SIMBAD database
as a K0III, and we derive a magnitude difference of approximately
2.5, which is slightly higher at bluer wavelengths. This suggests that
while the primary is evolved, the secondary is still near the main
sequence, and so a study could be done along the lines of 
Davidson \ea (2009) to determine the age and individual masses of the
two stars from the photometry through isochrone fitting.
Also, the orbit we calculate has very high eccentricity,
which helps explain why there were no successful observations of the
pair in the late 1990's: periastron passage occurred in 1997, with 
predicted separation of 6 mas.

\subsection{CHR 74}

A well-known speckle binary whose first interferometric observations date
from the CHARA speckle program in the mid-1980's, this 
object went unobserved from 1996 until the present work. The new data
suggest that nearly half an orbit has been completed since the 
first speckle observation. The primary is probably close to an A0V
star, and with a magnitude difference of perhaps 2 to 2.5, we estimate
that the secondary is an F2V star. This would suggest a mass sum in
the range of 4 to 5 $M_{\odot}$, which is lower than what is implied
by the value of $P$ and $a$ we derive, together with the {\it Hipparcos}
parallax of $5.04 \pm 0.39$ mas. We expect that the next decade will
provide an opportunity to improve upon the orbit presented here since
the separation will be above 0.1 arcsec and the position angle is expected
to cover about 30 degrees. We encourage other observers to put this object
back on their observing lists for the next few years.

\subsection{YSC 133}

A first orbit for this system was reported recently in Tokovinin \ea (2015),
but those authors assumed a quadrant flip of one data point taken by 
Horch \ea (2008). That was not an unreasonable assumption based on the 
relatively small magnitude difference, but the points presented here give
support to an ``alternate'' orbit where no quadrant flips are needed and
the period is about half of that derived in the earlier calculation. Given the 
revised {\it Hipparcos} parallax of $\pi = 13.55 \pm 0.43$ mas, the 
new orbit gives a mass sum of $3.5 \pm 1.0$ solar masses, which is 
somewhat high for a system of composite spectral type F7, but with
substantial uncertainty at this stage. In contrast, 
the Tokovinin orbit gives a mass
sum of about 1 solar mass, which would seem to be too low by roughly
the same amount.

\subsection{HDS 2947}

The new observations of this object presented in Table 3 indicate that
the orbit is eccentric, and that we observed the system near periastron
(and below the diffraction limit of the DCT). The orbital elements
obtained confirm that picture, yielding a total mass of $3.4 \pm 0.5$
$M_{\odot}$. The composite spectral type is F5 and the system has a
magnitude difference of about 1, so that we might suggest that the 
individual spectral types are F4 and F9. If so, then the total mass 
would be on the order of 2.5 $M_{\odot}$, which is about 2-$\sigma$ 
lower than the 
value implied by the current orbital elements and the parallax of 
$\pi = 13.96 \pm 0.55$ mas. However, because there is only one 
point near periastron at the moment, the orbital elements will of
course be modified with time.

\subsection{YSC 139}

The data at hand suggest that this F8 pair with small magnitude
difference is viewed nearly edge-on. The mass implied by the
spectral type is perhaps 2-2.5 $M_{\odot}$, if the system were
to have solar metallicity, again lower than that
implied by the orbital elements. The metallicity of the system
is -0.31 according to the Geneva-Copenhagen Catalog (Nordstr\"{o}m
\ea 2007), which would tend to lower the mass at a given spectral
type, thus making the discrepancy worse. Nonetheless, given the
large uncertainty in the dynamical mass, it is no worse than a 
2-$\sigma$ difference at this point. The mass ratio in the 
Geneva-Copenhagen catalog is 0.9, so this at least is consistent with the 
small magnitude difference obtained from speckle measures.

\section{Summary}

We have presented 1146 observations of double stars and suspected double
stars where the data were obtained 
with the Differential Speckle Survey Instrument
at Lowell Observatory's Discovery Channel Telescope.
938 of these yielded relative astrometry measures of 
high quality, with individual uncertainties in separation being
approximately 1.8 mas, and for most objects, uncertainties in 
position angle are less than 0.5 degrees. In the remaining 208 observations
presented, we found no companion to the limit of our detection
ability, and we have stated a 5-$\sigma$ detection limit at 0.2 arc seconds
in these cases.
Orbital parameters were calculated for six systems in the hope that they
will be observed by other groups over the next few years in
order to refine the orbital data for eventual use in astrophysical
studies of these stars.

\acknowledgments

The authors would like to thank all of the excellent staff at Lowell
Observatory for their help during our observing runs. 
We used the SIMBAD database, 
the Washington Double Star Catalog, the Fourth
Catalog of Interferometric Measures of Binary Stars, and the Sixth Orbit
Catalog in the preparation of this paper.
This work is 
funded by NSF grant 1429015. 
J.W.D. acknowledges support from a Small Grant through the Faculty 
Development Committee at Albion College.
L.A.C. acknowledges support from the Foundation for Undergraduate 
Research, Scholarship, and Creative Activity program at Albion College.

% References -----------------------------------------------------------

\clearpage

% Tables -----------------------------

\begin{centering}
\begin{deluxetable}{lcccc}
\tabletypesize{\scriptsize}
\tablewidth{0pt}
\tablenum{1}
\tablecaption{Final Values for the Pixel Scale and Orientation}
\tablehead{ 
\colhead{Run} &
\colhead{Channel A} &
\colhead{Channel A} &
\colhead{Channel B} &
\colhead{Channel B} \\

& \colhead{Offset Angle} &
\colhead{Scale} &
\colhead{Offset Angle} &
\colhead{Scale} 

} 
\startdata 
Mar & $5.45^{\circ}$ & 18.433 mas/pix & $-5.77^{\circ}$ & 20.486 mas/pix \\
Jun & $5.27^{\circ}$ & 18.249 mas/pix & $-5.62^{\circ}$ & 20.345 mas/pix \\
Sep/Oct & $5.39^{\circ}$ & 18.781 mas/pix & $-5.39^{\circ}$ & 20.029 mas/pix \\
Nov & $5.39^{\circ}$ & 18.781 mas/pix & $-5.39^{\circ}$ & 20.029 mas/pix \\
\enddata 
\end{deluxetable}

\begin{deluxetable}{lclrcrrrrl}
\tabletypesize{\scriptsize}
\tablewidth{0pt}
\tablenum{2}
\tablecaption{Orbits and Residuals Used in the Scale Determination}
\tablehead{ 
\colhead{Run} &
\colhead{WDS} &
\colhead{Discoverer} &
\colhead{HIP} &
\colhead{Besselian} &
\colhead{$\Delta \theta_{A}$} &
\colhead{$\Delta \rho_{A}$} &
\colhead{$\Delta \theta_{B}$} &
\colhead{$\Delta \rho_{B}$} &
\colhead{Orbit Reference} \\
&& \colhead{Designation} &&
\colhead{Year} &
\colhead{($^{\circ}$)} &
\colhead{(mas)} &
\colhead{($^{\circ}$)} &
\colhead{(mas)} &
}

\startdata
Mar & $15232+3017$ & STF 1937AB & 75312 & 
      2014.2189 & $-$0.3 & $+$3.0 & $-$0.2 & $+$1.2 & Muterspaugh \ea 2010b\\
&&&&  2014.2244 & $+$0.2 & $-$1.7 & 0.0    & $+$1.3 & \\
    & $15278+2906$ & JEF 1      & 75695 & 
      2014.2189 & $+$0.3 & $+$1.7 & $+$0.4 & $+$1.5 & Muterspaugh \ea 2010b\\
&&&&  2014.2244 & $+$0.5 & $-$3.3 & $+$0.2 & $-$5.3 & \\
    & $17080+3556$ & HU 1176AB  & 83838 & 
      2014.2190 & $-$0.6 & $+$0.2 & $-$0.5 & 0.0    & Muterspaugh \ea 2010b\\

Jun & $15232+3017$ & STF 1937AB & 75312 & 
      2014.4594 & $-$0.2 & $-$0.3 & 0.0    & $-$0.5 & Muterspaugh \ea 2010b\\
    & $15278+2906$ & JEF 1      & 75695 & 
      2014.4594 & $-$0.1 & $+$0.1 & 0.0    & $+$0.5 & Muterspaugh \ea 2010b\\

Sep/Oct & $04136+0743$ & A 1938    & 19719 &
      2014.7584 &    0.0 & $+$0.2 &    0.0 & $+$0.2 & Muterspaugh \ea 2010b\\ 
        & $19490+1909$ & AGC 11AB  & 97496 &
      2014.7577 &    0.0 & $-$0.1 & $-$0.1 & $+$0.7 & Muterspaugh \ea 2010b\\ 
        & $21145+1000$ & STT 535AB & 104858 &
      2014.7552 & $+$0.4 & $+$0.2 & $+$0.4 & $+$0.5 & Muterspaugh \ea 2008\\ 
        & $22409+1433$ & HO 296AB  & 111974 &
      2014.7608 & $-$0.1 & $-$1.9 & $-$0.1 & $-$5.7 & Muterspaugh \ea 2010b\\ 
&&&&  2014.7636 & $-$0.3 & $+$1.8 & $-$0.2 & $+$1.5 & \\ 

\enddata 
\end{deluxetable}

%---insert table of measures here
%\documentclass[preprint]{aastex}
\pagestyle{empty}
%\begin{document}
\begin{deluxetable}{lllrrrrccl}
\tabletypesize{\scriptsize}
\tablenum{3}
\tablecaption{DCT double star speckle measures}
\tablehead{ 
\colhead{WDS} & HR,ADS &
\colhead{Discoverer} & HIP &  
\colhead{Date} & 
 \colhead{$\theta$} & \colhead{$\rho$} & 
\colhead{$\Delta m$} & 
\colhead{$\lambda$} & \colhead{$\Delta \lambda$} \\
\colhead{($\alpha$,$\delta$ J2000.0)} & \colhead{HD, or DM} & 
\colhead{Designation} & &
\colhead{($2000+$)} & 
\colhead{($^{\circ}$)} & 
\colhead{(${\prime \prime}$)} &
\colhead{(mag)} & 
\colhead{(nm)} & \colhead{(nm)} 
} 
\startdata

 $00012+1358$ & BD$+$13 5195  & LSC 2 & 96  
 & 2014.7580 & 184.5 & 0.2168 &  2.35 & 692 & \hspace{1pt} 40  \\ 
&&&
 & 2014.7580 & 184.6 & 0.2166 &  1.59 & 880 & \hspace{1pt} 50  \\ 

 $00067+0839$ & ...  & LSC 3 & 551  
 & 2014.7580 &  15.9 & 0.5730 &  4.07 & 692 & \hspace{1pt} 40  \\ 
&&&
 & 2014.7580 &  15.7 & 0.5755 &  3.16 & 880 & \hspace{1pt} 50  \\ 

 $00087-0213$ & HD 406 & LSC 4 & 700  
 & 2014.7553 & 109.0 & 0.9242 &   ... & 692 & \hspace{1pt} 40  \\ 
&&&
 & 2014.7553 & 108.7 & 0.9210 &   ... & 880 & \hspace{1pt} 50  \\ 

 $00121+5337$ & ADS 148 & BU 1026AB & 981  
 & 2014.7581 & 318.6 & 0.3436 &  1.02 & 692 & \hspace{1pt} 40  \\ 
&&&
 & 2014.7581 & 318.6 & 0.3447 &  0.93 & 880 & \hspace{1pt} 50  \\
\enddata 
 
\tablenotetext{a}{Quadrant ambiguous.}
\tablenotetext{b}{Quadrant inconsistent with previous measures in the 4th 
Interferometric Catalog of Hartkopf \ea 2001b.}
\tablenotetext{c}{Third component noted, but above 1.2 arcsec.}
\tablenotetext{d}{This object was used in the determination of the pixel scale and
detector orientation, so only the magnitude difference measure is reported here.}

\end{deluxetable} 
%\end{document} 

\begin{deluxetable}{clrcrrrrl}
\tabletypesize{\scriptsize}
\tablewidth{0pt}
\tablenum{4}
\tablecaption{Ephemeris Positions and Residuals Used in the Astrometric Accuracy Study}
\tablehead{ 
%\colhead{Coord} &
\colhead{WDS} &
\colhead{Discoverer} &
\colhead{HIP} &
\colhead{Besselian} &
\colhead{$\theta_{eph}$} &
\colhead{$\rho_{eph}$} &
\colhead{${\Delta \bar{\theta}}$} &
\colhead{${\Delta \bar{\rho}}$} &
\colhead{Orbit Reference} \\
& \colhead{Designation} &&
\colhead{Year} &
\colhead{($^{\circ}$)} &
\colhead{($^{\prime \prime}$)} &
\colhead{($^{\circ}$)} &
\colhead{(mas)} &
}

\startdata
$00121+5337$ & BU 1026AB &    981 & 2014.7581 & 
  $319.1 \pm 1.8$ & $0.3474 \pm 0.0088$ & -0.5 & (-3.3) & Hartkopf \ea 1996 \\
$00507+6415$ & MCA 2 &   3951 & 2014.7554 & 
  $136.6 \pm 9.8$ & $0.0477 \pm 0.0019$ & (-3.6) & -8.0 & Mason \ea 1997 \\
$01057+2128$ & YR 6Aa,Ab &   5131 & 2014.7555 & 
  $297.9 \pm 2.2$ & $0.1057 \pm 0.0064$ & -0.3 & (-1.4) & Horch \ea 2010 \\
$01072+3839$ & A 1516AB &   5249 & 2014.7554 & 
  $24.9 \pm 4.4$  & $0.1479 \pm 0.0024$ & (+4.6) & +5.4 & Hartkopf \ea 2000 \\
$01083+5455$ & WCK 1Aa,Ab &   5336 & 2014.7581 & 
  $42.3 \pm 1.6$  & $1.1127 \pm 0.0475$ &  -4.1  & (-41.0) & Drummond \ea 1995 \\
$02022+3643$ & A 1813AB &   9500 & 2014.7582 & 
  $130.2 \pm 12.4$ & $0.0393 \pm 0.0036$ & (+2.3) & +0.7 & Hartkopf \ea 2000 \\
$02157+2503$ & COU 79 &  10535 & 2014.7582 & 
  $38.6 \pm 0.1$ & $0.1933 \pm  0.0014$ &  +0.6 & +2.0 & Muterspaugh \ea 2010b \\
$02396-1152$ & FIN 312 &  12390 & 2014.7557 & 
  $87.0 \pm 0.3$ & $0.0864 \pm  0.0004$ &  +0.4 &  -3.6 & Docobo \& Andrade 2013 \\
\enddata
\tablenotetext{a}{Although the separation uncertainty meets the criteria
discussed in the text, this object will probably need an orbit refinement
after passing through periastron, as discussed in Horch \ea 2015, 
so we have not used the separations here.
This does not affect the position angles, as it is an edge-on system.}
\end{deluxetable}

%---insert table of non-detections here
%\documentclass[preprint]{aastex}
%\pagestyle{empty}
%\begin{document}
\begin{deluxetable}{lrcccl}
\tabletypesize{\scriptsize}
\tablewidth{0pt}
\tablenum{5}
\tablecaption{High-Quality Nondetections and 5-$\sigma$ Detection Limits}
\tablehead{
\colhead{($\alpha$,$\delta$ J2000.0)} & HIP &
\colhead{Date} &
\colhead{5-$\sigma$ Det. Lim.} &
\colhead{5-$\sigma$ Det. Lim.} & Notes\tablenotemark{a} \\
\colhead{(WDS format)} &  &
\colhead{($2000+$)} &
\colhead{(692 nm)} &
\colhead{(880 nm)}
}
\startdata
01320$+$5712 &   7141 & 14.7581 & 4.43 & 3.61 & YSC 25; previous $\Delta m$=4.92 (698 nm) \\
01473$+$0659 &   8319 & 14.7555 & 3.32 & 3.27 &  OCC 46; {\it Hipparcos} suspected double %/YSCU
\\
02282$+$2940 &  11486 & 14.7582 & 4.45 & 4.33 & YSC 1; previous sep of 90 mas (2003) \\
02330$+$1910 &  11855 & 14.7557 & 3.47 & 3.80 & {\it Hipparcos} suspected double%/YSCU 
\\
02366$+$1227 &  12153 & 14.7557 & 3.18 & 4.14 & MCA 7; G-C spectroscopic binary\tablenotemark{b}; \\
           &        &         &      &      & ephemeris sep. near diff. lim. \\
03233$-$0748 &  15776 & 14.7556 & 3.74 & 3.79 & Probable thick disk member. \\
03234$+$3059 &  15781 & 14.7557 & 4.01 & 4.03 & {\it Hipparcos} suspected double%/YSCU 
\\
\enddata
\tablenotetext{a}{If previous observations are mentioned, these are drawn from
the 4th Interferometric Catalogue of Binary Stars (Hartkopf \ea 2001b).}
\tablenotetext{b}{The notation ``G-C'' refers to the Geneva-Copenhagen spectroscopic
survey of Nordstr\"{o}m \ea 2007.}
\end{deluxetable}
%\end{document}

\clearpage

\begin{deluxetable}{lcccccc}
\tabletypesize{\scriptsize}
\tablewidth{0pt}
\tablenum{6}
\tablecaption{Preliminary Orbits for Six Systems}
\tablehead{ 
\colhead{Parameter} &
\colhead{EGG 2Aa,Ab} & \colhead{HDS 2143} & \colhead{CHR 74} &
\colhead{YSC 133} & \colhead{HDS 2947} & \colhead{YSC 139} 
} 
\startdata 
HIP                & 10772                & 74649             & 91118            &  91880            & 102029             & 116360 \\
Spectrum           & F2                   & K0III             & A0V              &  F7V              & F5                 & F8 \\
$\Delta m$         & 1.0                  & 2.5               & 2.2              &  0.6              & 1.0                & 0.4 \\
$P$, years         & $29.71 \pm 0.07$     & $66 \pm 5$        & $68.4  \pm 1.6$  &  $7.32 \pm 0.14$  & $18.1 \pm 0.3$     & $3.64 \pm 0.05$ \\
$a$, mas           & $136.0 \pm 0.9$      & $229 \pm 2$       & $168.4 \pm 1.5$  &  $78 \pm 7$       & $145 \pm 4$        & $69 \pm 6$ \\
$i$, degrees       & $7. \pm 6.$          & $132 \pm 4$       & $116.6 \pm 0.6$  &  $108 \pm 2$      & $104.3 \pm 1.4$    & $87.9 \pm 0.8$ \\
$\Omega$, degrees  & $166. \pm 3.$        & $354 \pm 8$       & $40.2 \pm 0.9$   &  $120 \pm 5$      & $99.9 \pm 1.4$     & $90 \pm 2$ \\
$T_{0}$, years     & $2013.02 \pm 0.05$   & $2063 \pm 5$      & $2042.1 \pm 2.3$ &  $2009.7 \pm 0.4$ & $2014.81 \pm 0.05$ & $2013.41 \pm 0.05$ \\
\enddata 
\end{deluxetable}

\end{centering}

\clearpage

\begin{figure}[tb]
\plottwo{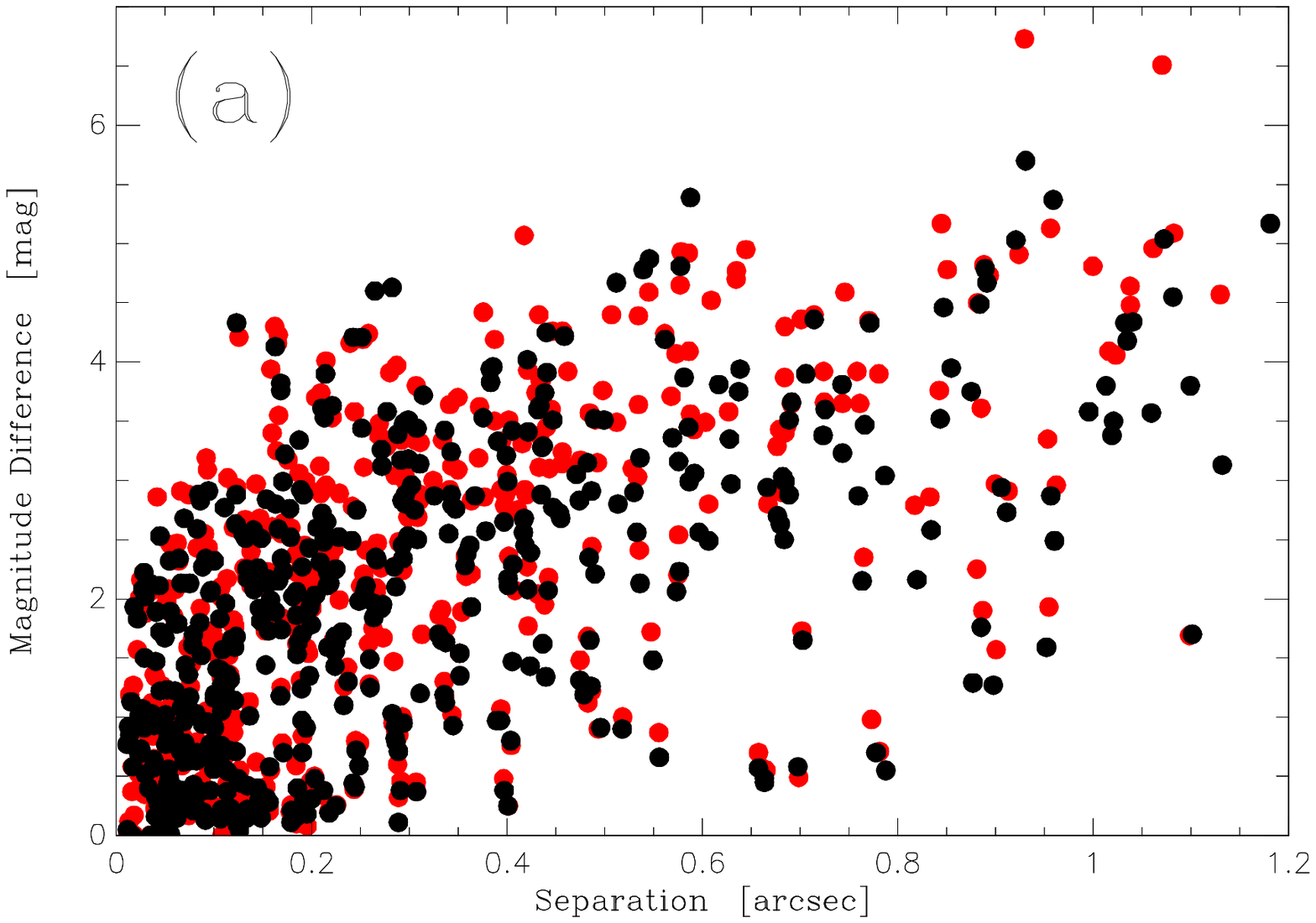}{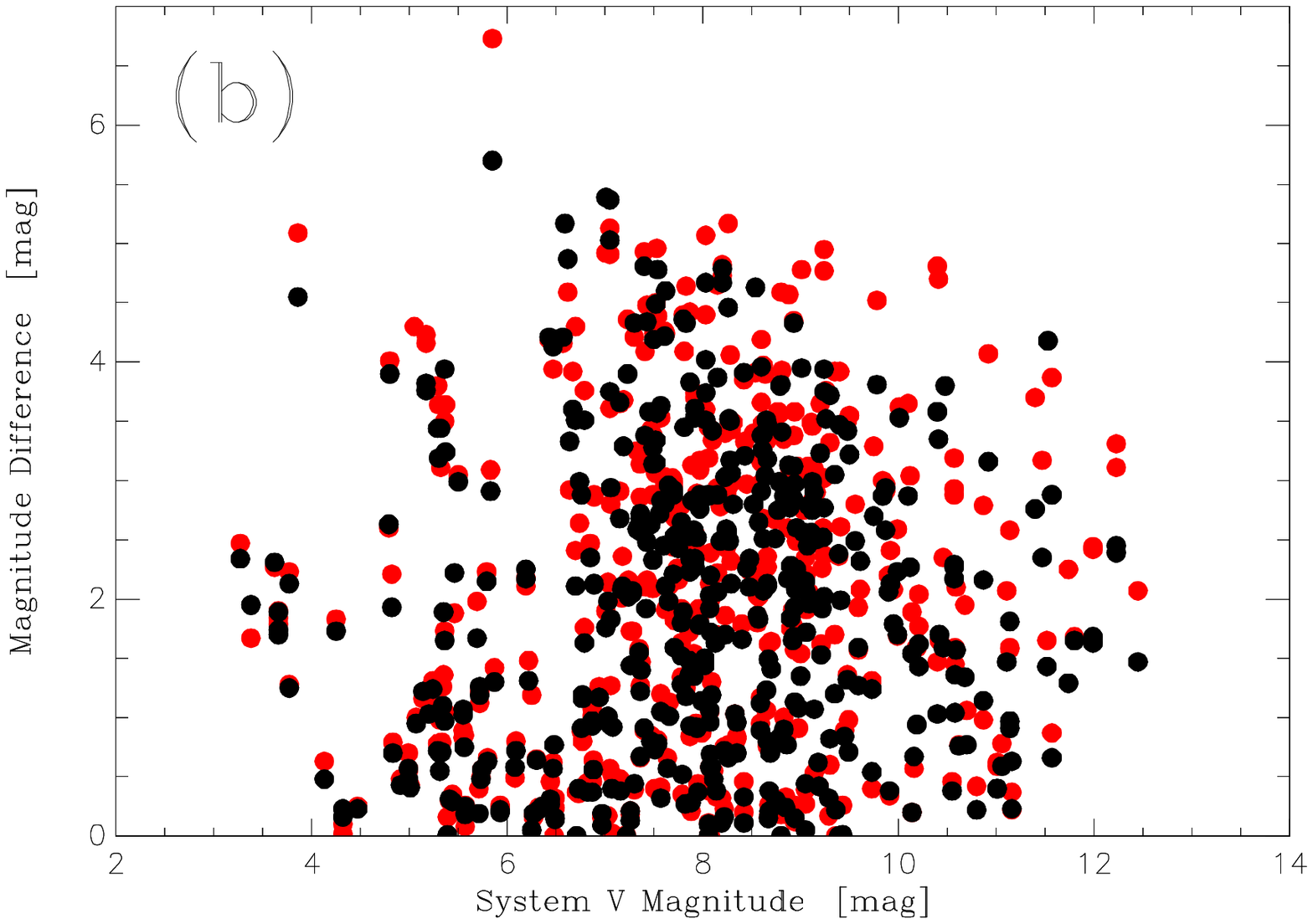}
\caption{
(a) Magnitude difference as a function of separation for the measures
listed in Table 3. 
(b) Magnitude difference as a function of system
$V$ magnitude for the measures listed in Table 3.
In both plots, the red circles are measures taken with the 692 nm filter
and black circles are measures in the 880 nm filter.
}
\end{figure}

\clearpage
                                                                                
\begin{figure}[tb]
\plottwo{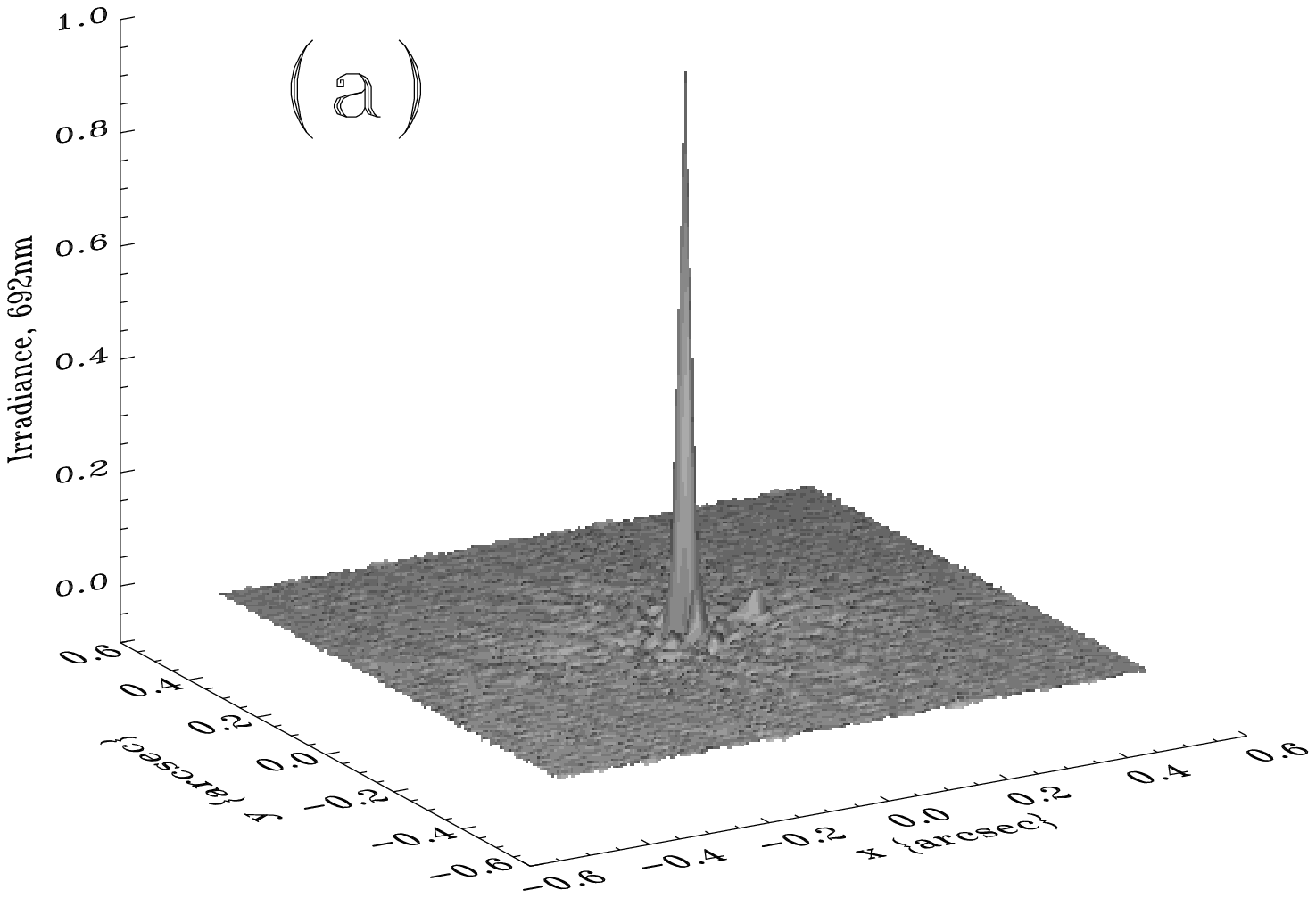}{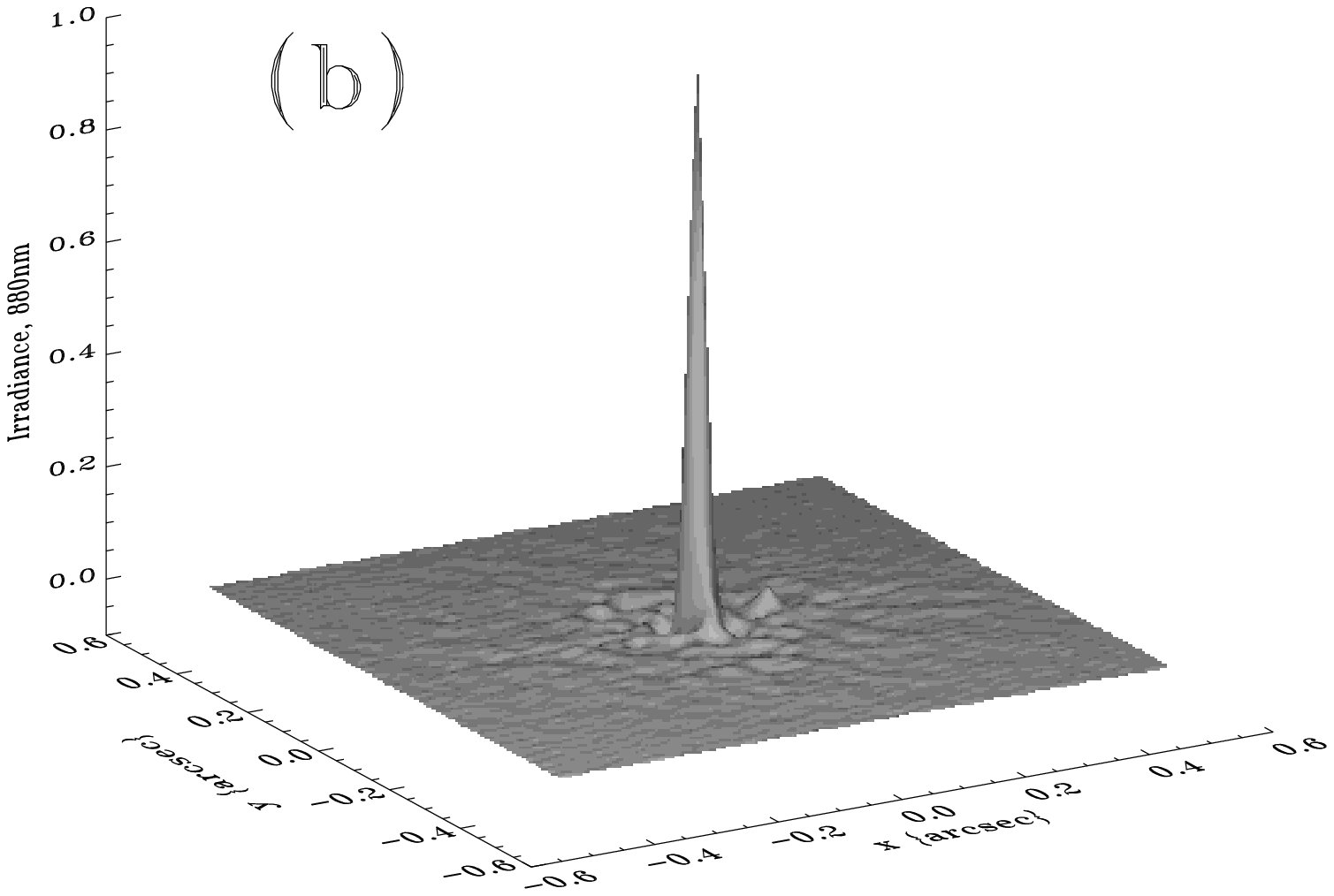}

\plottwo{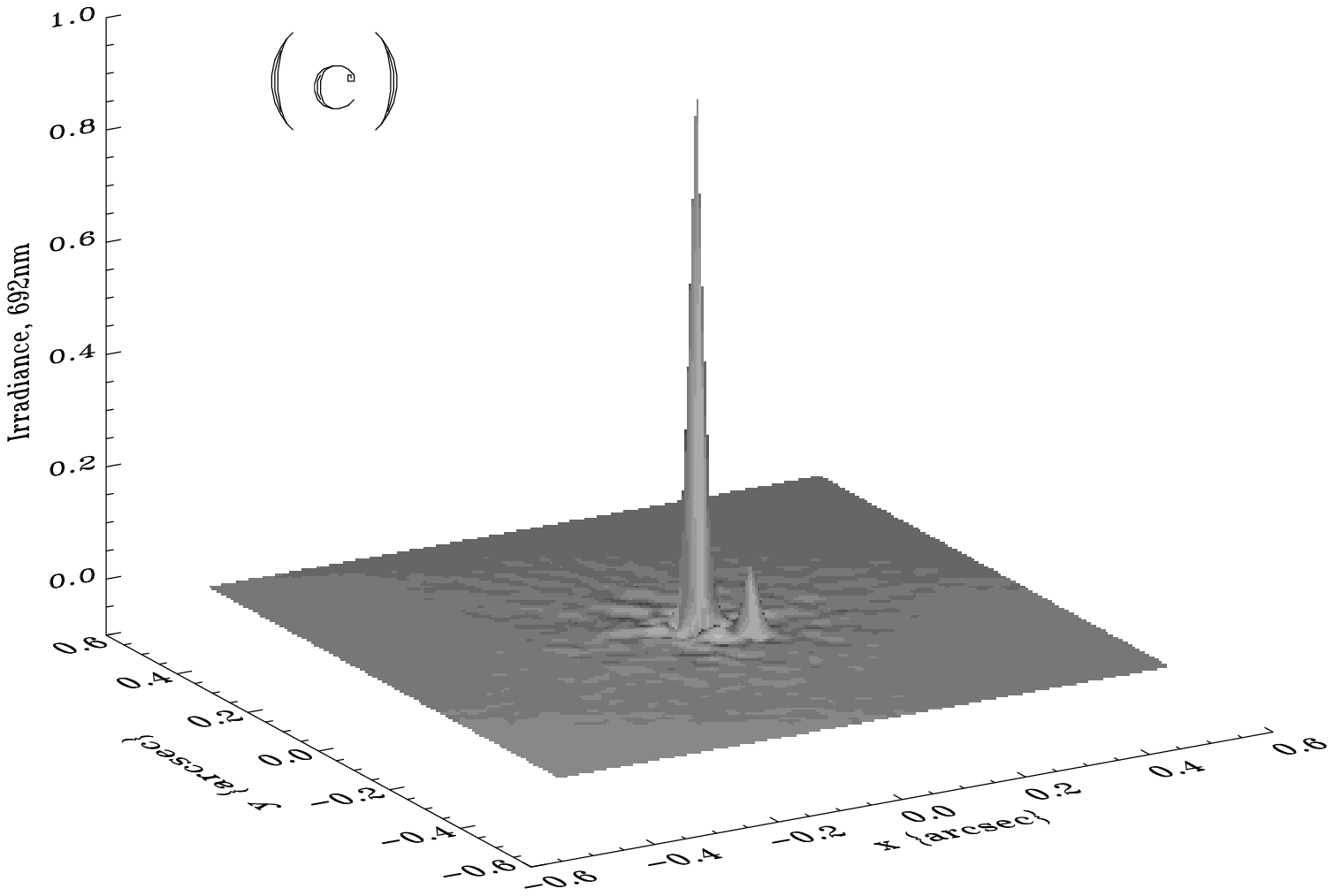}{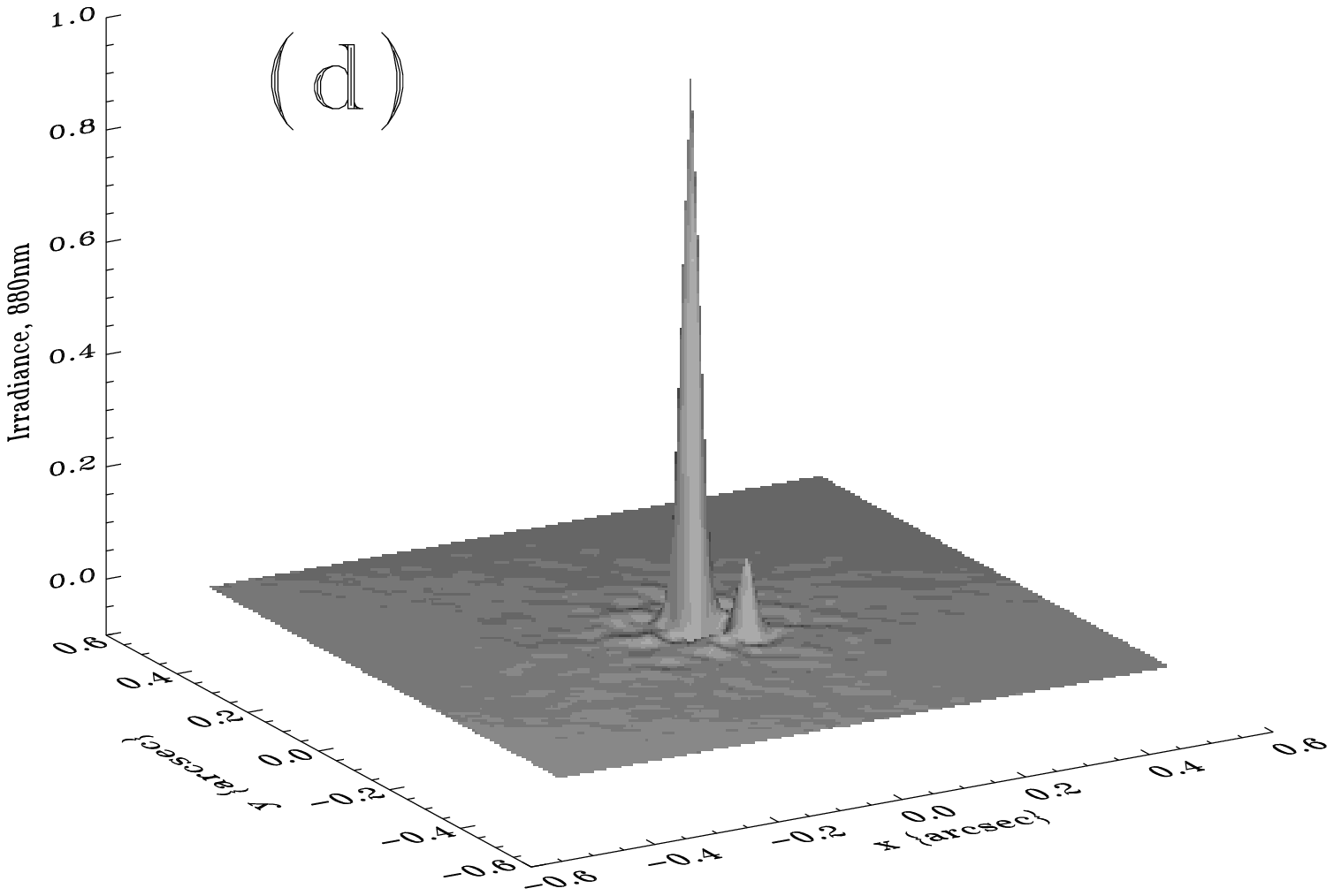}
\caption{
Reconstructed images of two targets. 
(a) HDS 2143 = HIP 74649 in the 692 nm filter. %hr5675=h074649=hds2143
(b) HDS 2143 = HIP 74649 in the 880 nm filter. %hr5675=h074649=hds2143
(c) CHR 74 = HIP 91118 in the 692 nm filter. %h091118
(d) CHR 74 = HIP 91118 in the 880 nm filter. %h091118
}
\end{figure}

\clearpage

\begin{figure}[tb]
\plottwo{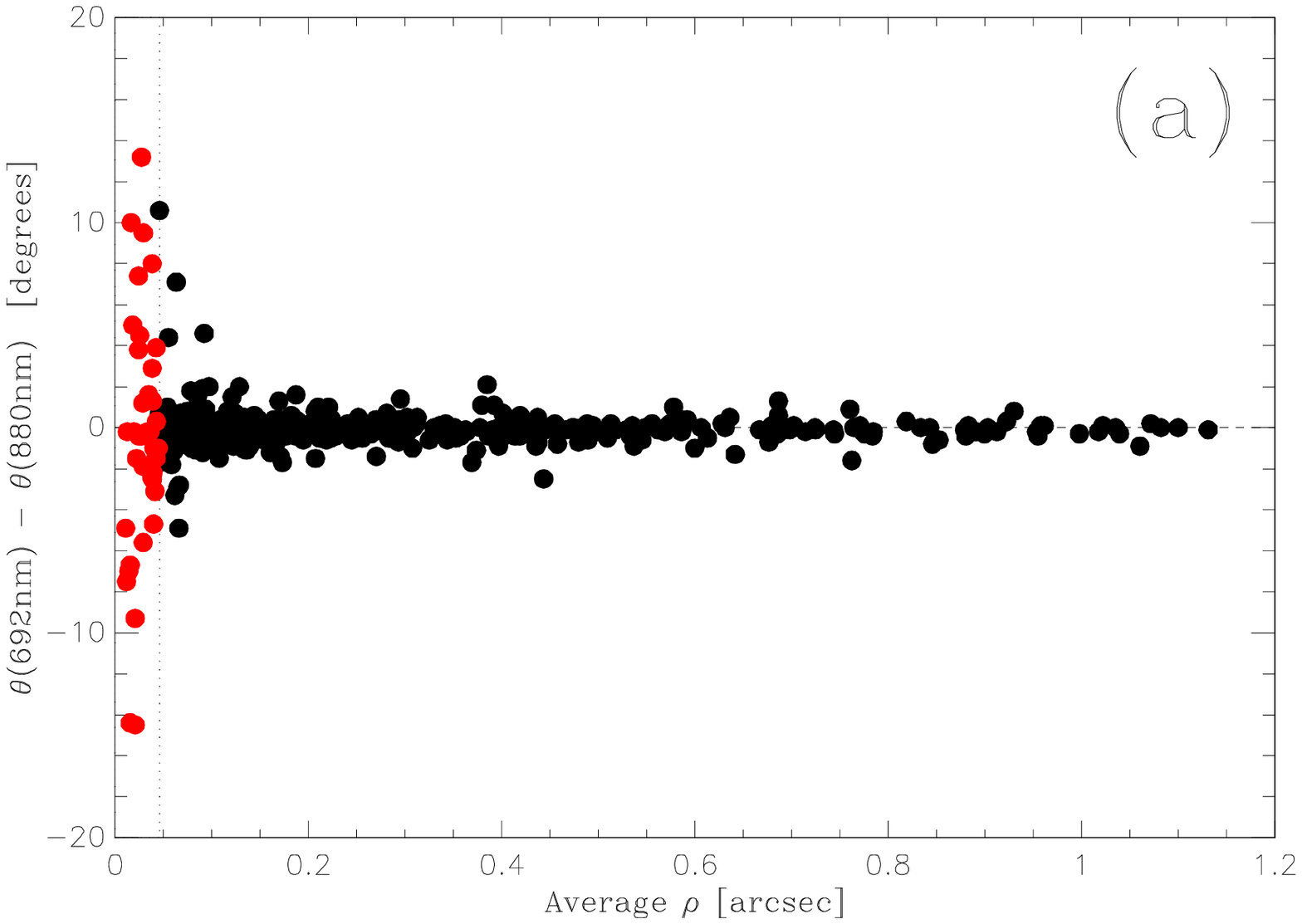}{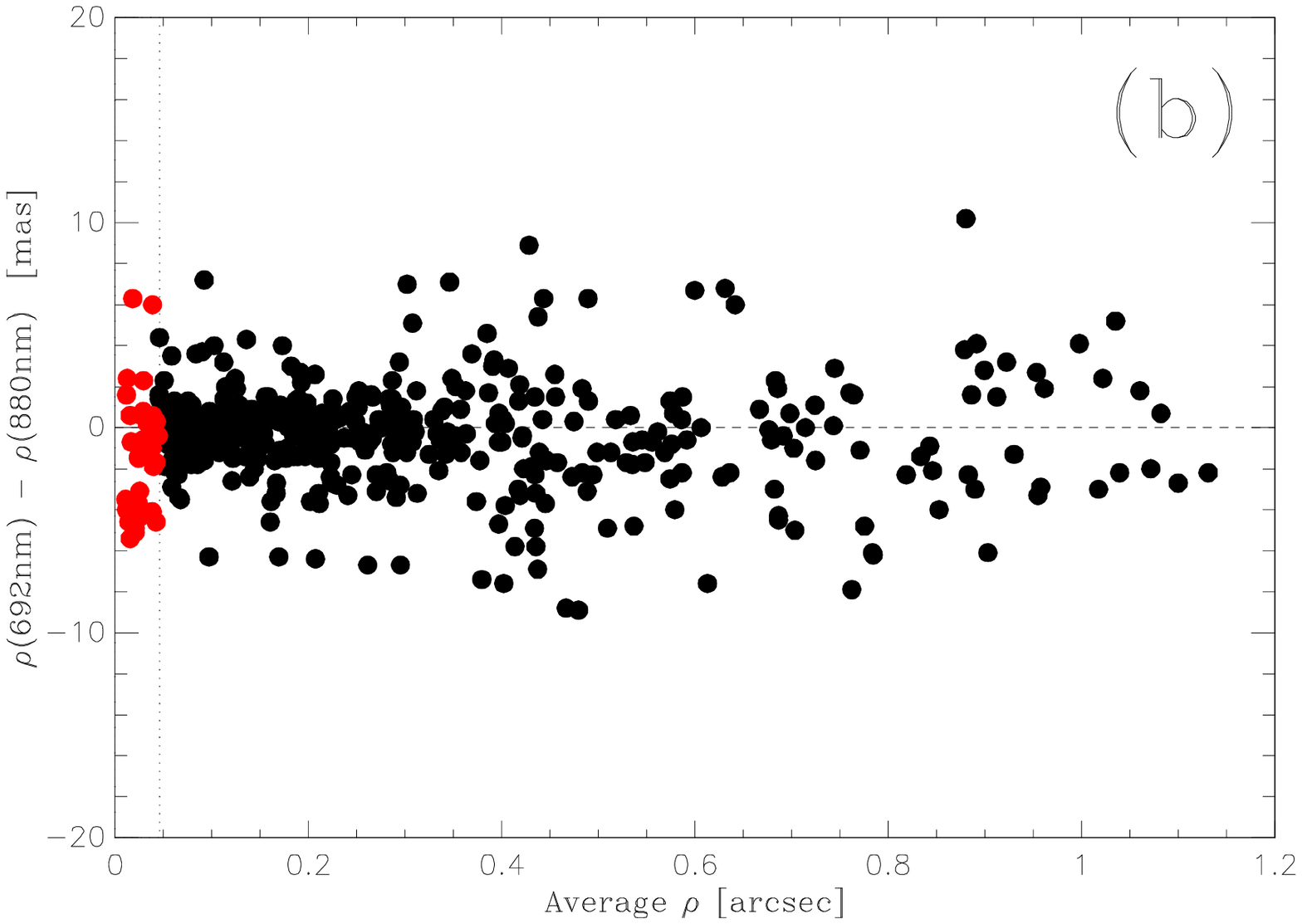}

\plottwo{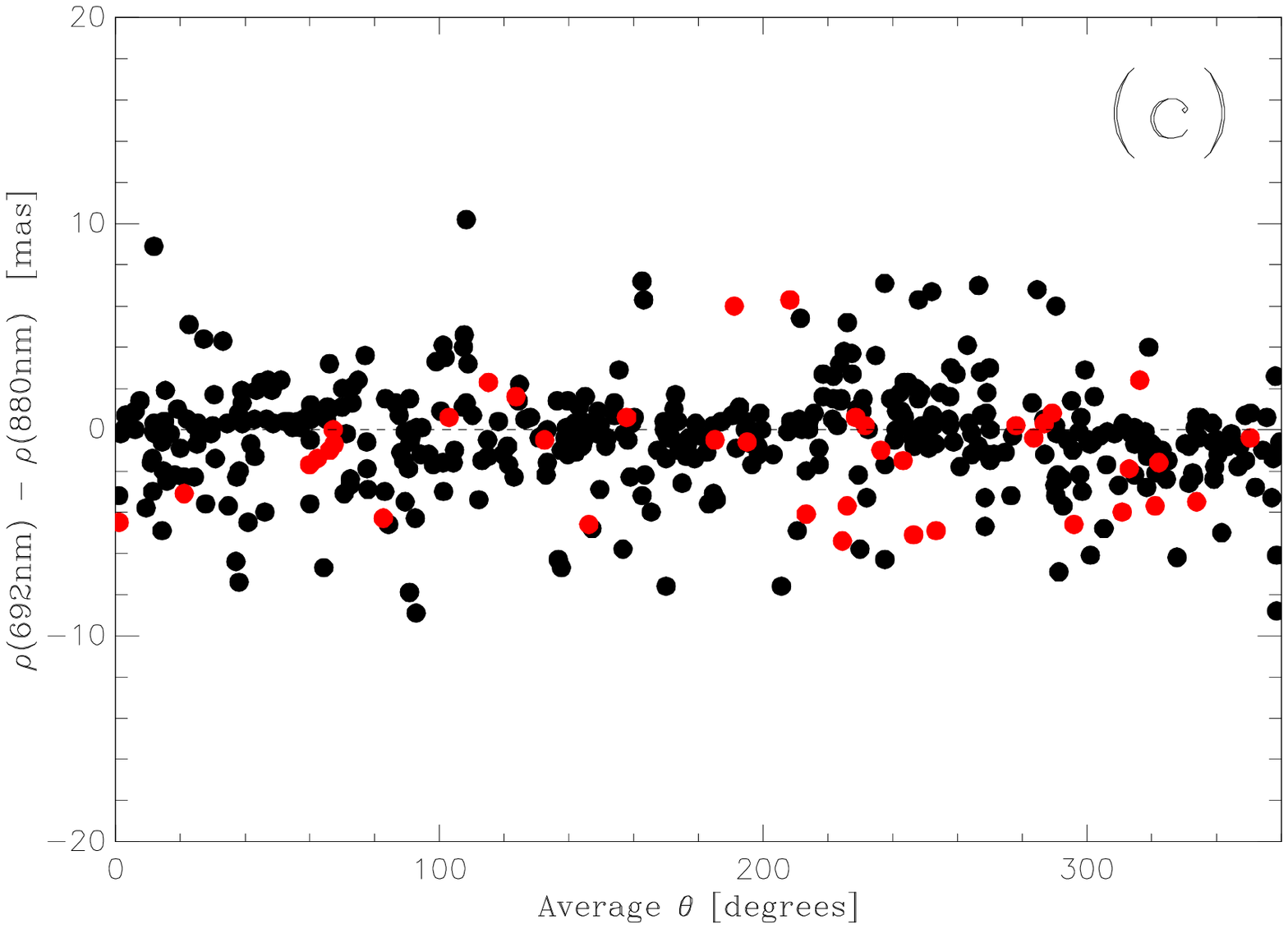}{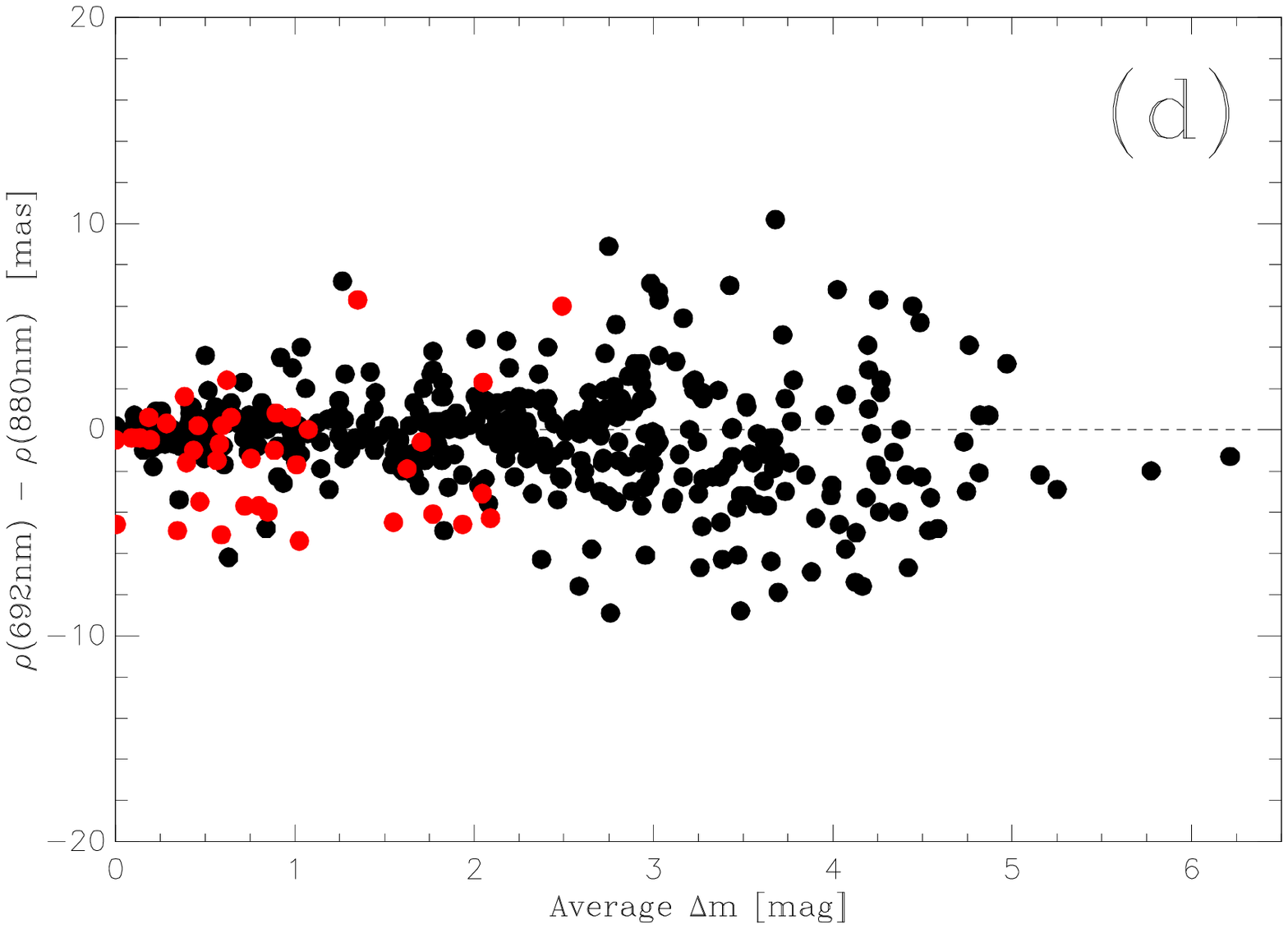}
\caption{
Measurement differences between the two channels of the instrument
plotted as a function of measured separation, position angle, or
magnitude difference.
(a) Position angle ($\theta$) differences as a function of average separation.
(b) Separation ($\rho$) differences as a function of average separation.
In both plots, the dotted vertical line at the left marks the average 
diffraction limit of the two wavelengths used, and squares indicate
measures below the diffraction limit.
(c) Separation differences as a function of average position angle.
(d) Separation differences as a function of average magnitude difference.
In all plots, the filled red circles 
indicate measures below the diffraction limit.
}
\end{figure}

\begin{figure}[tb]
\plottwo{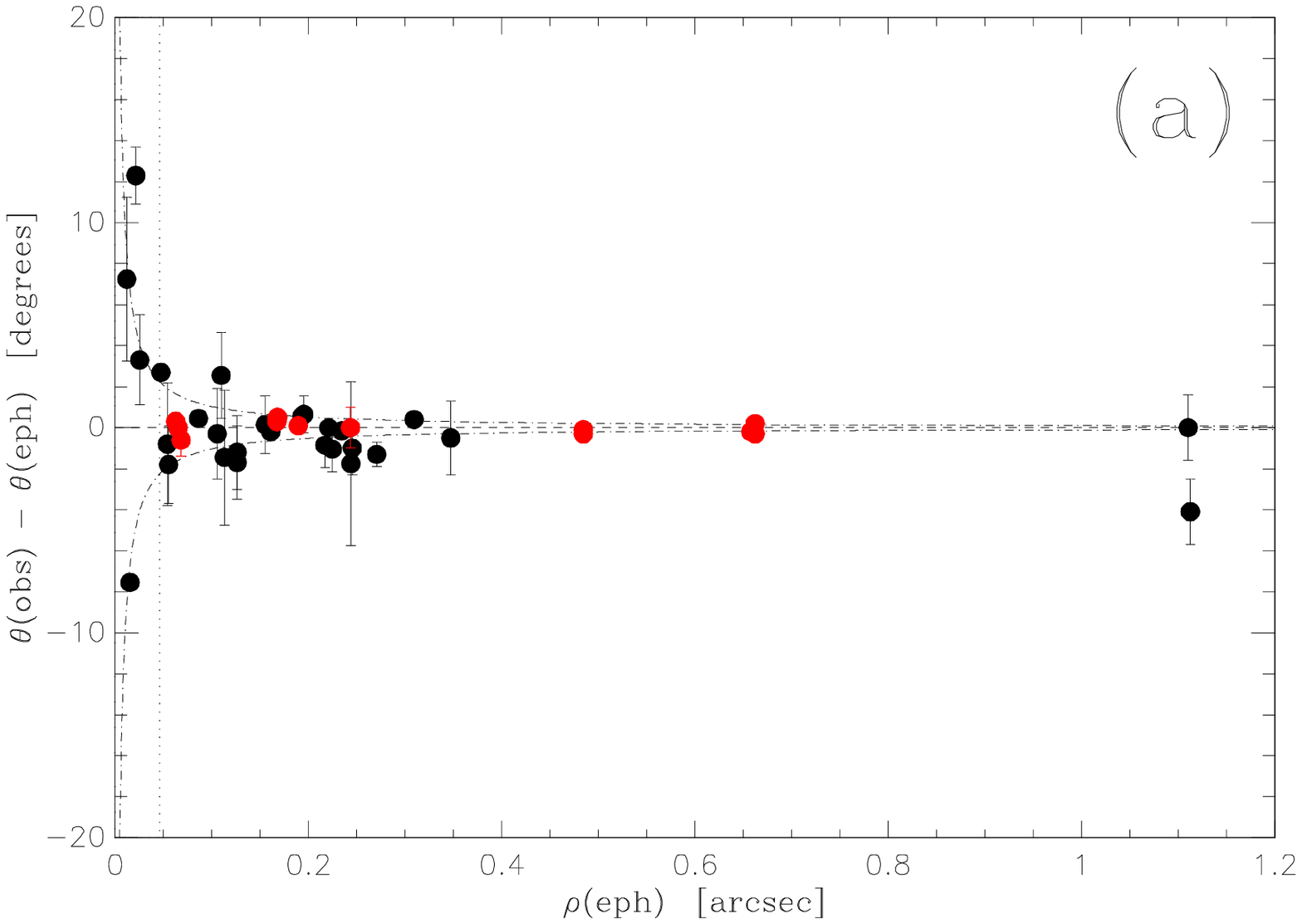}{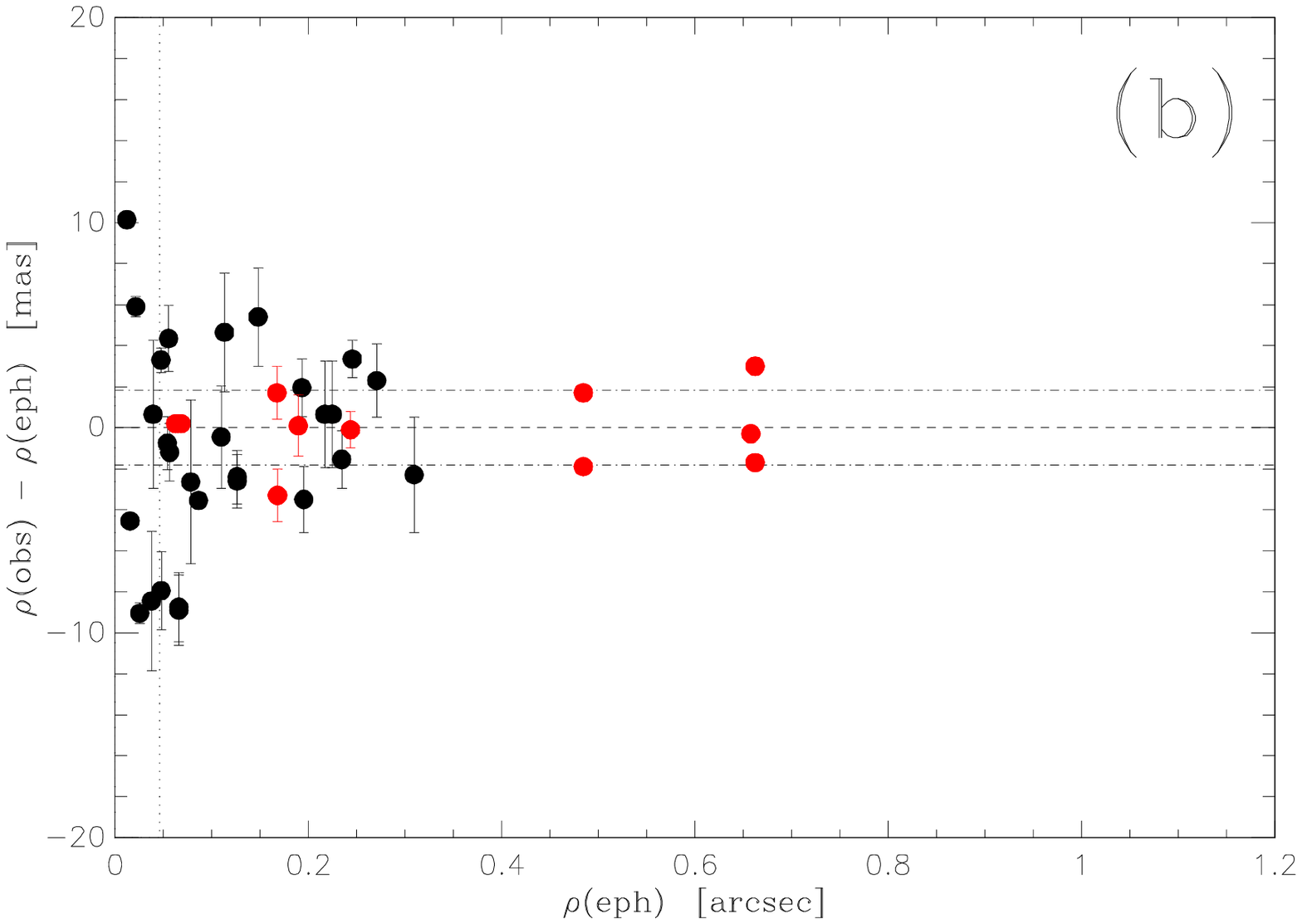}
\caption{
Observed minus ephemeris differences in position angle and separation
when comparing the measures presented here with orbital ephemerides
of objects having an orbit in the Sixth
Orbit Catalog of Hartkopf, Mason and Worley (2001) with the quality
criteria described in the text.
(a) Position angle residuals. In this plot, the dot-dash curves mark the
position angle error expected from a linear measurement error
of 1.82 mas, the derived value for a single-channel measurement.
(b) Separation residuals.
The dot-dash line is drawn at 1.82 mas. 
In both plots, the dotted vertical line at the left of
the plot marks the average 
of the diffraction limits in the 692 nm and 880 nm filters, and
the error bars indicate the uncertainties in the ephemeris position
based on error propagation of the published uncertainties in the 
orbital elements. 
Filled red circles indicate 
the observations used to determine the scale, and are not used in the
derivation of the statistics mentioned in the text for this study.
}
\end{figure}

\begin{figure}[tb]
\plottwo{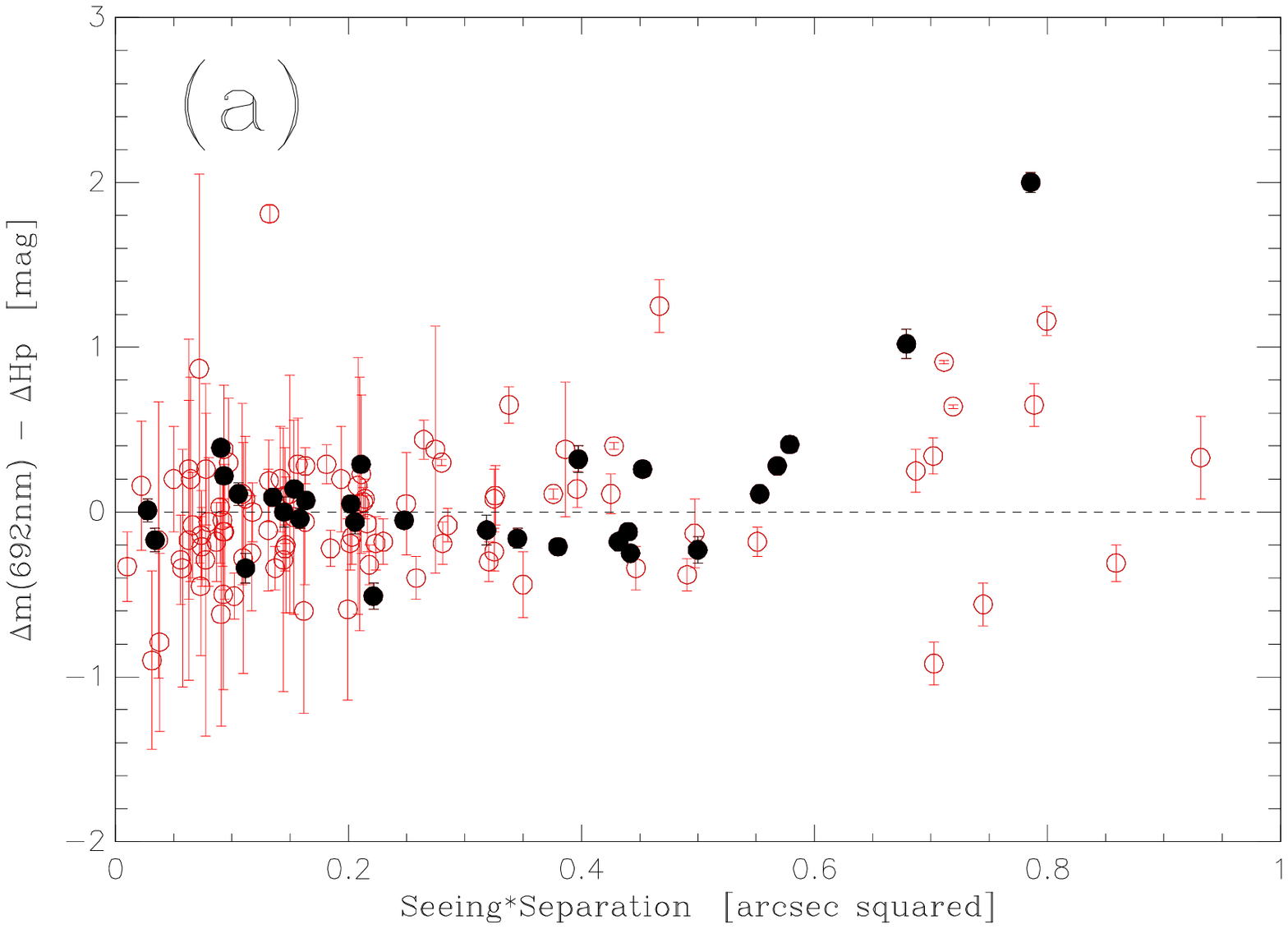}{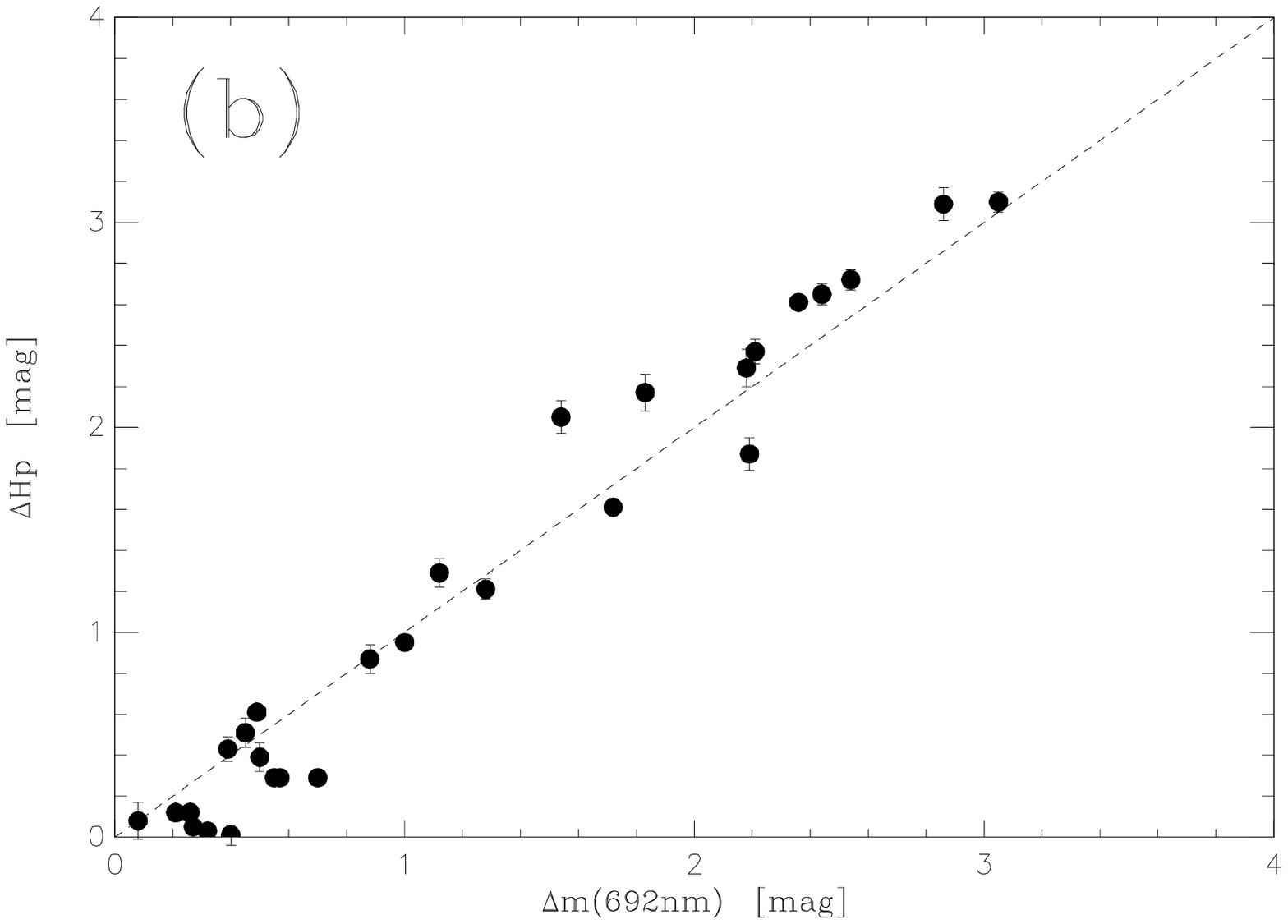}
\caption{
A comparison of the differential photometry presented in Table 1 with 
Hipparcos differential photometry.
(a) The difference in $\Delta m$ between our measure at 692 nm and the 
$\Delta H_{p}$ value appearing in the {\it Hipparcos} Catalogue as 
a function of the parameter $q^{\prime}$ = seeing times 
separation discussed in the text. Stars with indications of variability
and/or no $B-V$ value in the {\it Hipparcos} Catalogue are not considered.
Filled black circles indicate those systems with a $B-V < 0.6$ 
and an uncertainty in the $\Delta H_{p}$ value
of less than 0.10 magnitudes. Open red circles have no color cut and
no cut in $\Delta H_{p}$.
(b) A plot of the $\Delta H_{p}$ value as a function of the magnitude
difference in Table 1 for those systems with $B-V$ color 
less than $+$0.6 and $\delta(\Delta H_{p}) < 0.10$ mag. 
In both plots, the error bars are the uncertainties
appearing in the {\it Hipparcos} Catalogue. 
}
\end{figure}

\begin{figure}[tb]
\plottwo{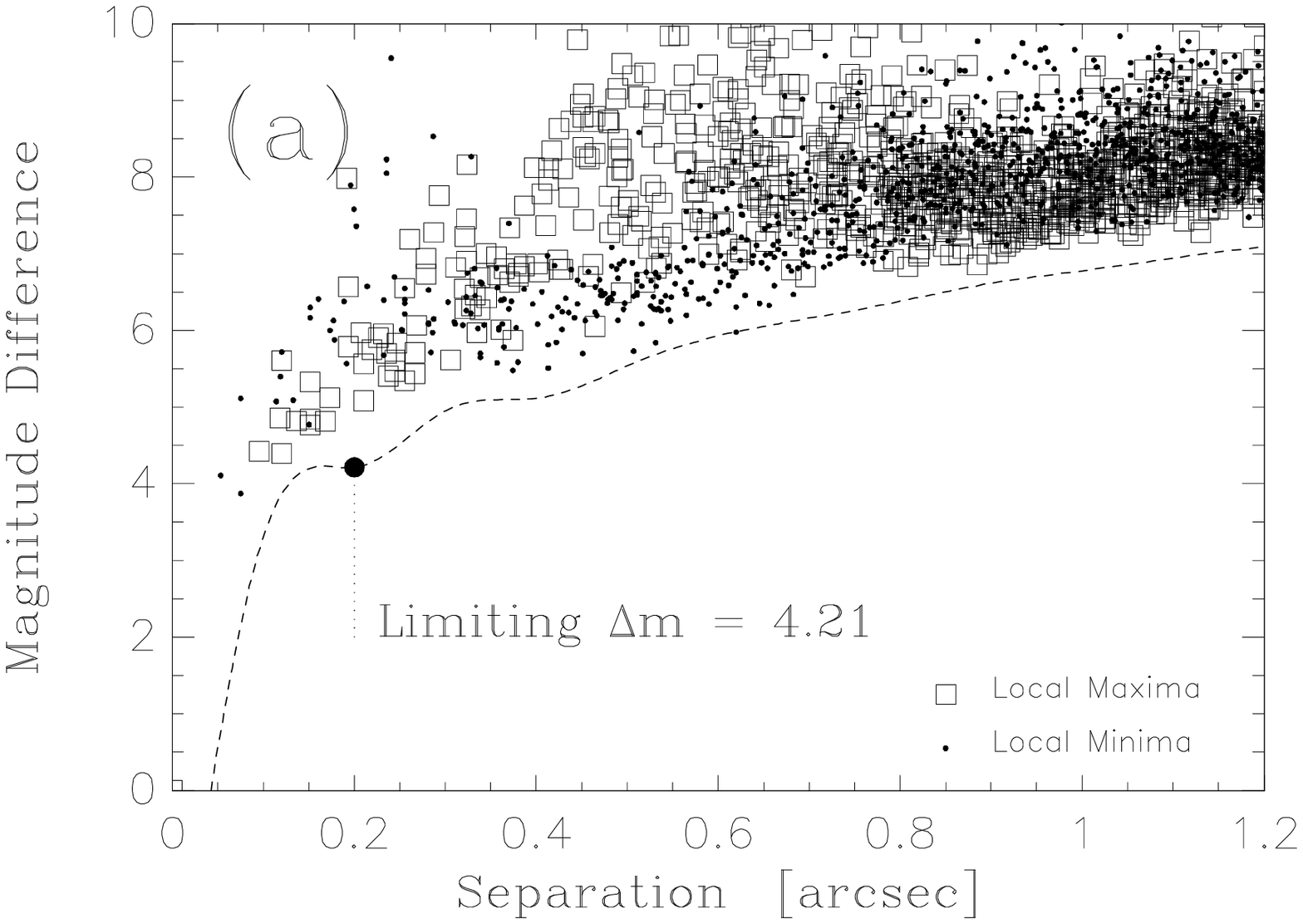}{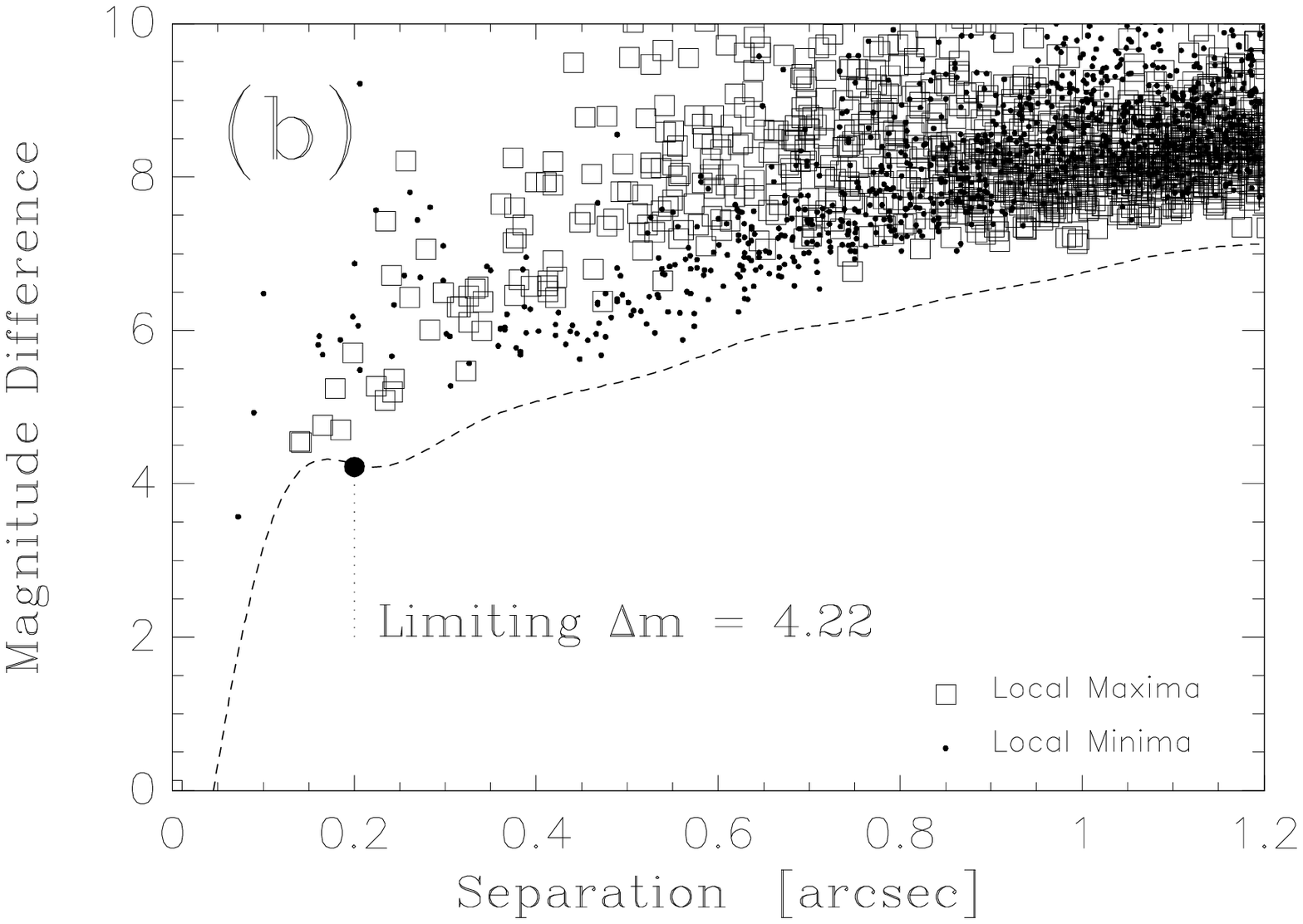}

\plottwo{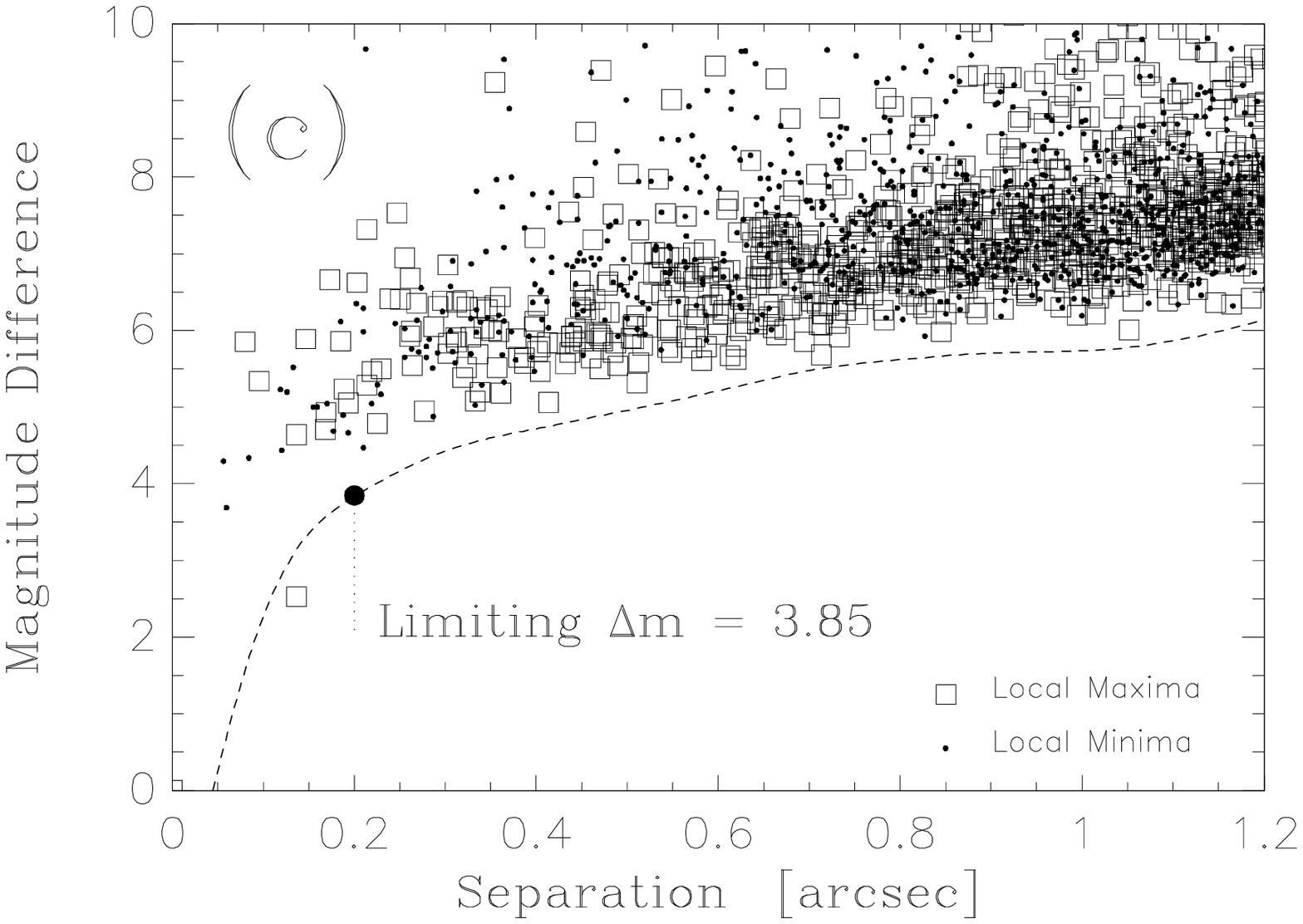}{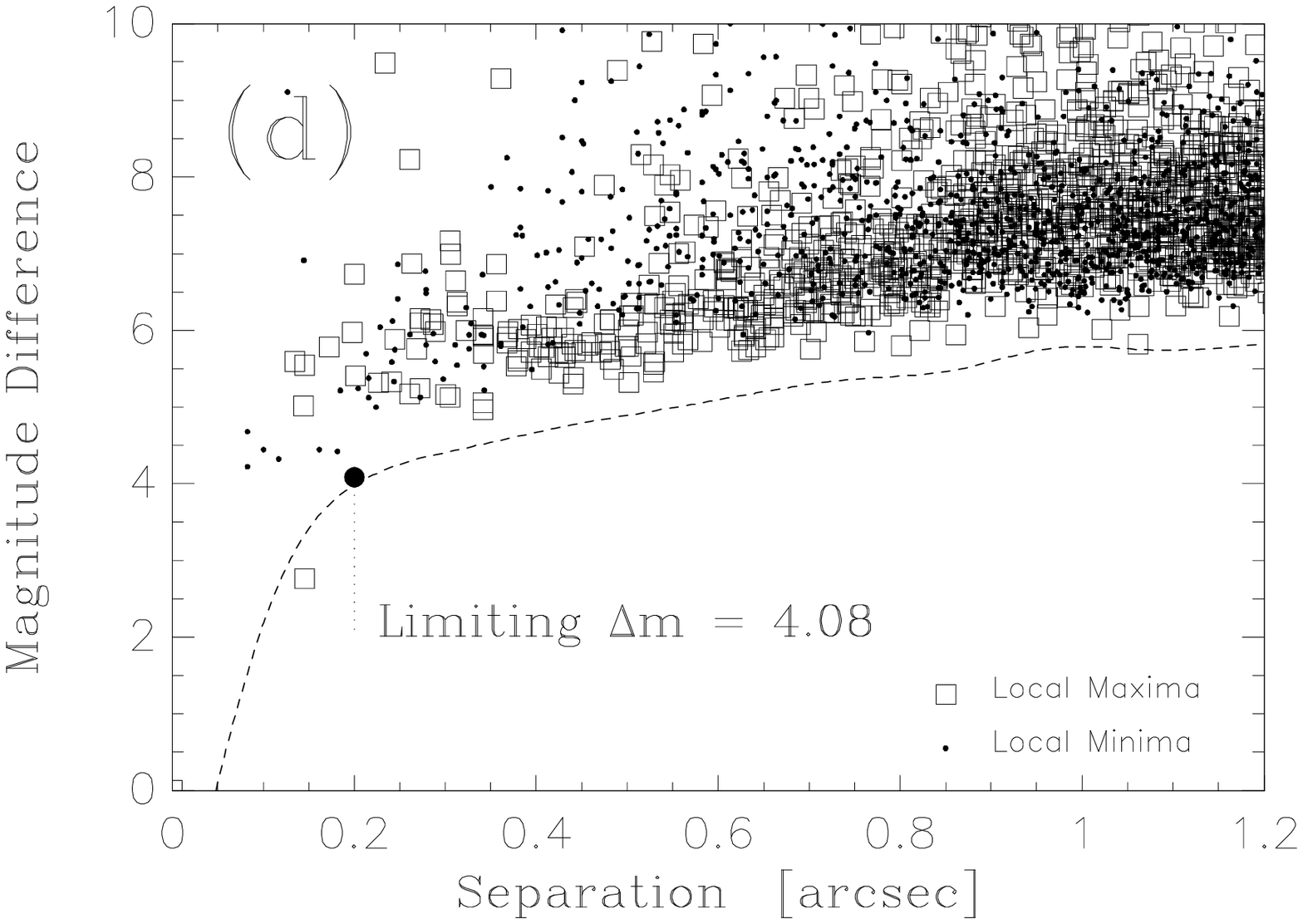}
\caption{
Detection limit analysis for HIP 77986 = 4 Her as described in the text.
(a) The result in the 692 nm filter.
(b) The result in the 880 nm filter.
For comparison, the detected double star HIP 113690 = LSC 104. 
(c) The result in the 692 nm filter.
(d) The result in the 880 nm filter.
Note that in both plots for this object, a secondary is clearly detected 
below the 5-$\sigma$ curve ({\it i.e.\ }it is more than a 5-$\sigma$ result
statistically) 
at a separation of approximately 0.13 arc 
seconds.
}
\end{figure}

\begin{figure}[tb]
\plottwo{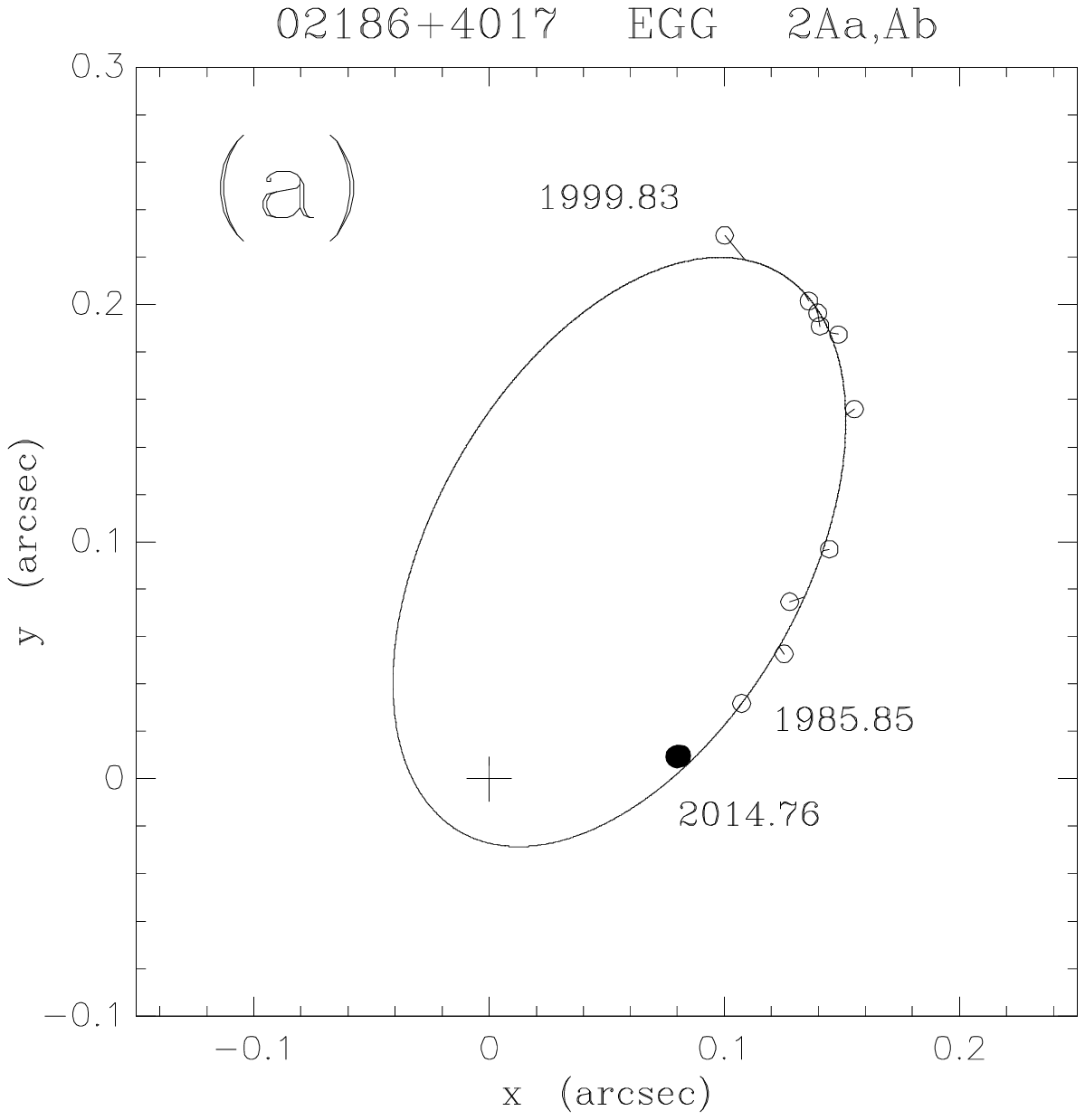}{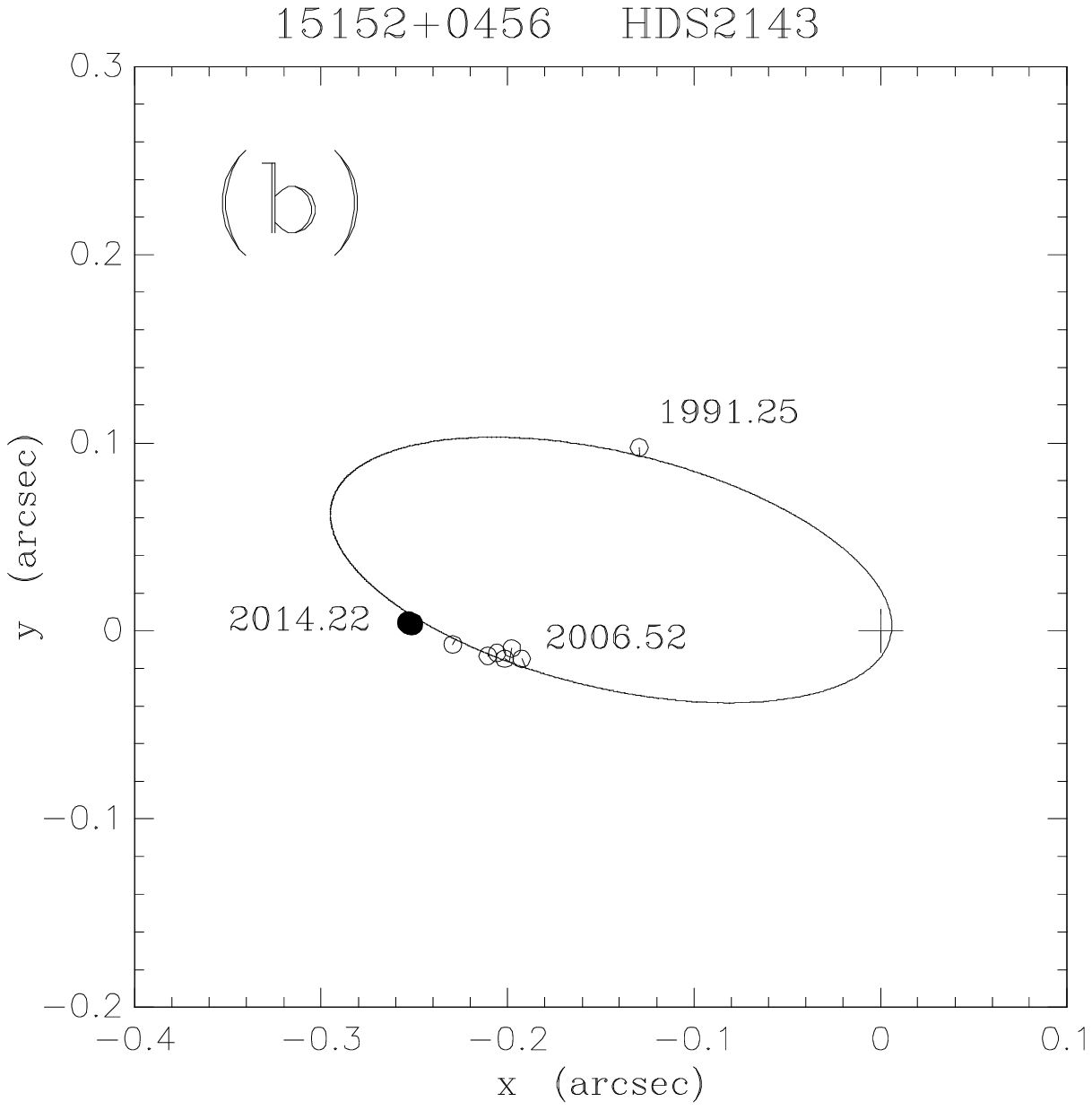}

\plottwo{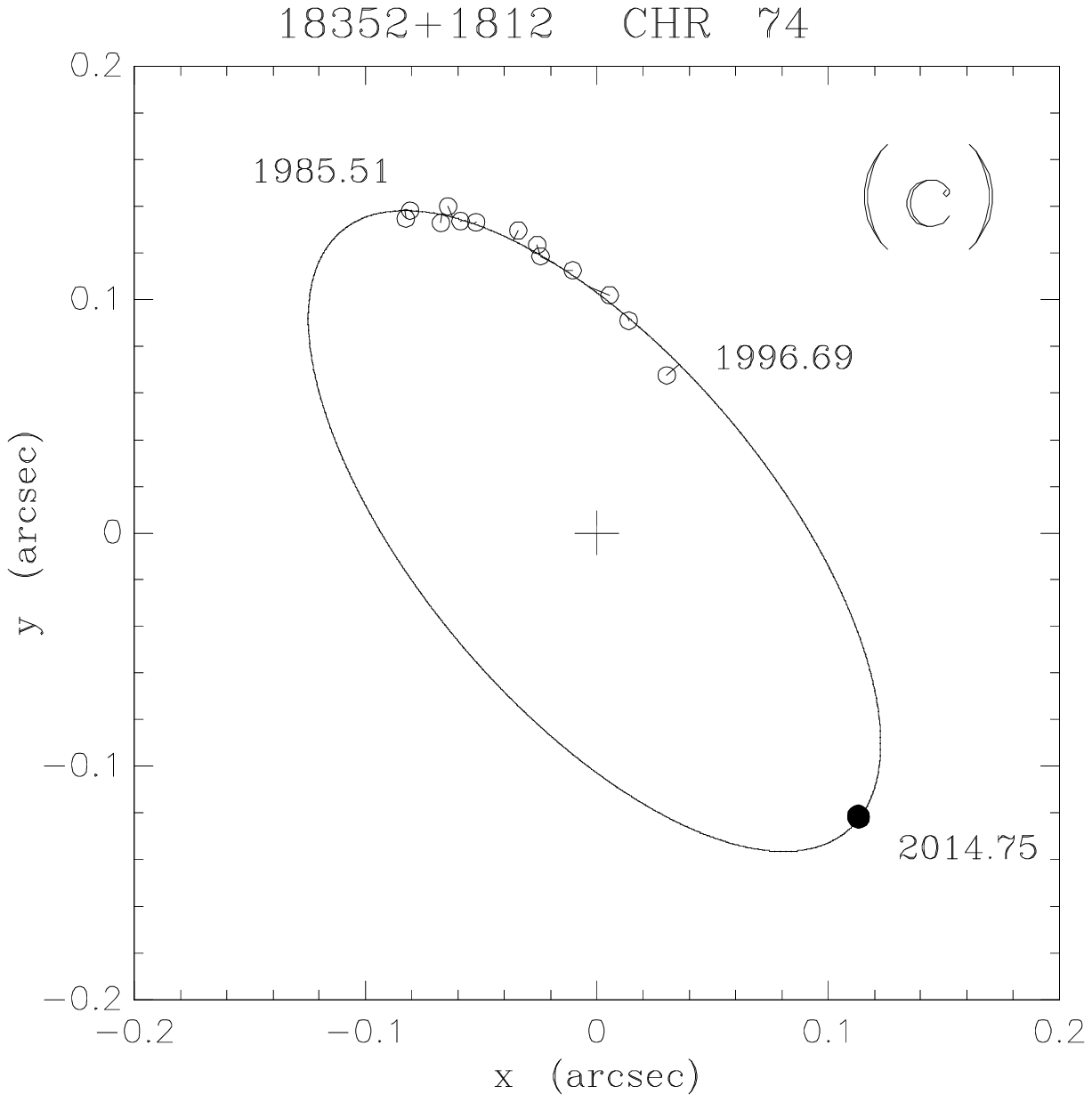}{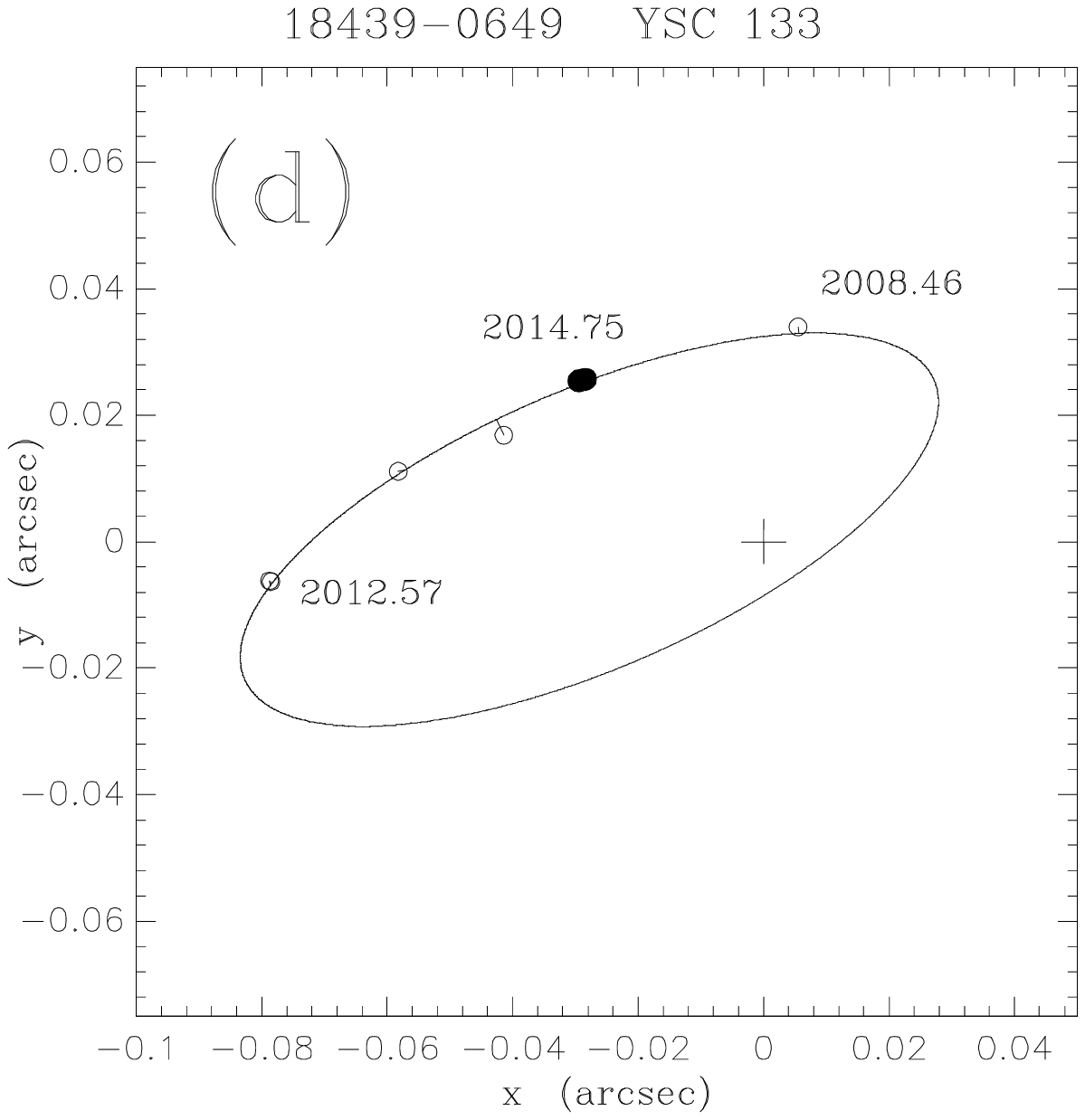}

\plottwo{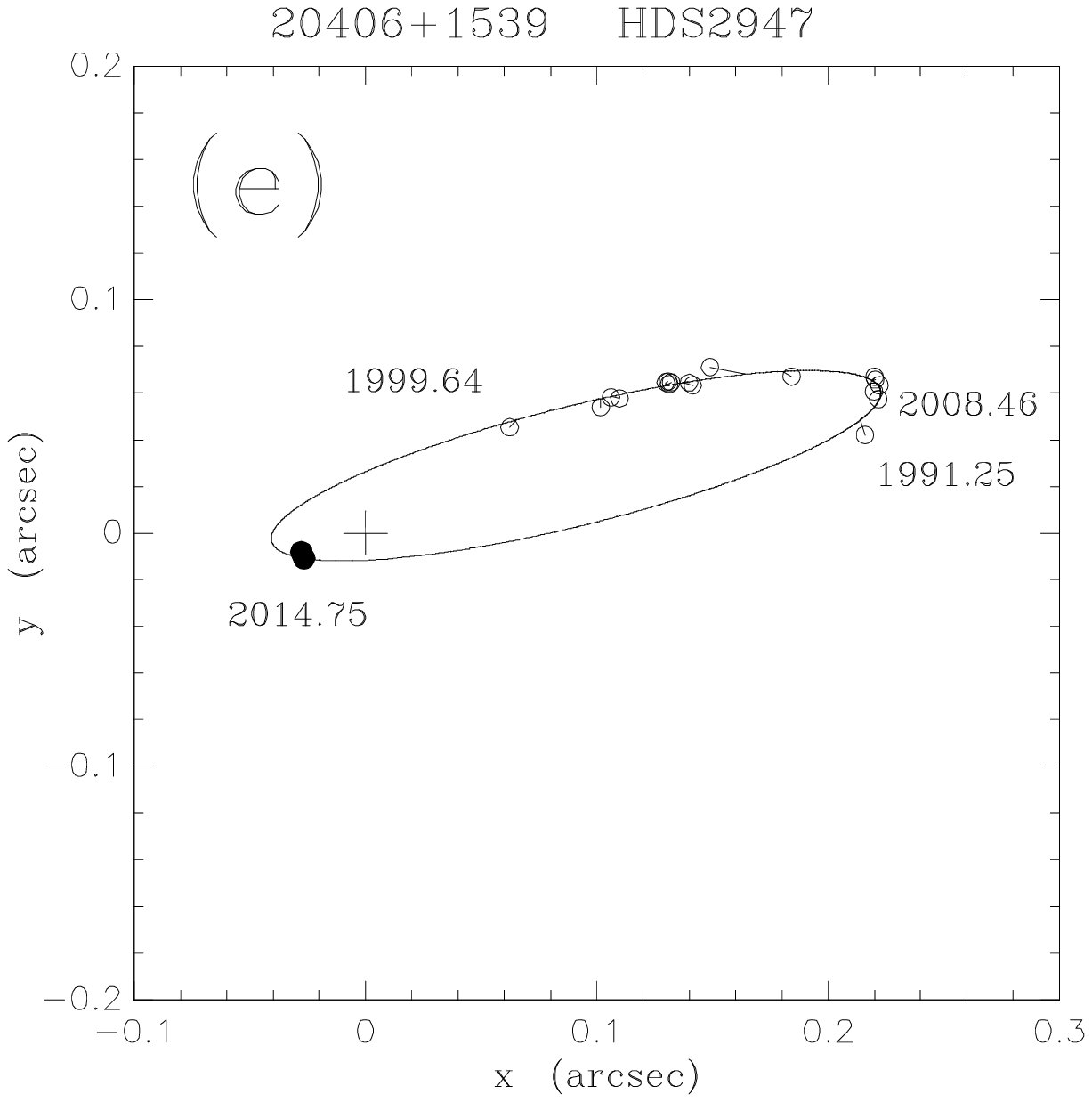}{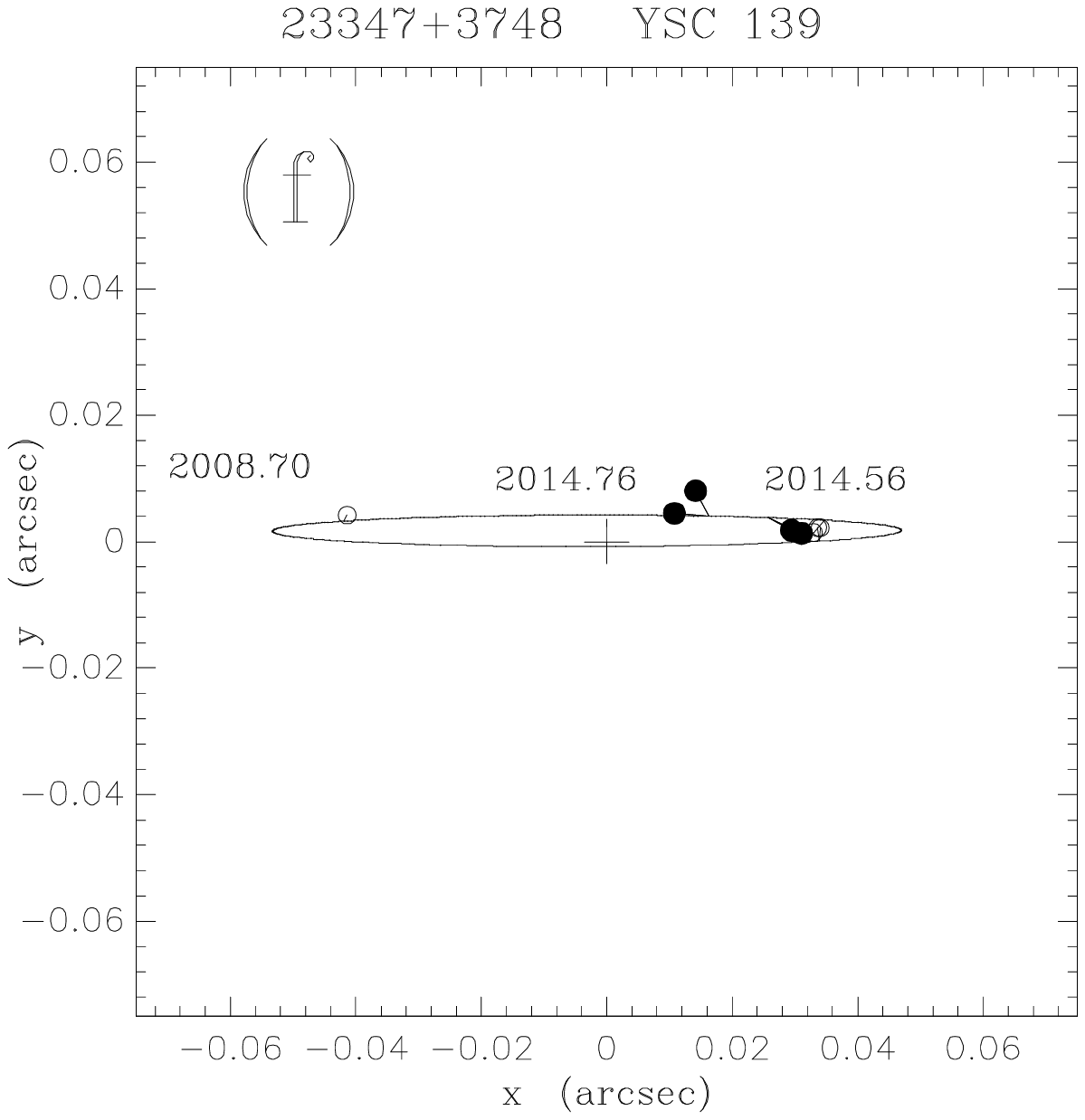}
\caption{
Orbits for the systems listed in Table 6, using data in the literature
as well as our points in Table 3, which are shown as the filled circles.
All points are drawn with line segments from the data point to the 
location of the ephemeris prediction on the orbital path.
North is down and East is to the right. 
}
\end{figure}

\end{document}